\documentclass{IEEEtran}

\newif\iftr %
\trtrue 
\trfalse     
\newif\ifshort 
\shorttrue     
\newif\ifnony  
\nonytrue     
\newif\ifccs   
\ccstrue      

\usepackage[usenames]{xcolor} 
\definecolor{darkblue}{rgb}{0,0.08,0.45}
\colorlet{tr_colour}{orange} 

\usepackage[disable]{todonotes}
\newcommand{\bob}[1]{\todo[color=olive!40,inline]{Bob: #1}}
\newcommand{\bobk}[1]{\todo[color=green!40,inline]{Bob for Koen \& Olivier: #1}}

\usepackage{ifpdf}

\ifpdf
    \usepackage{graphicx}
    \usepackage[colorlinks,linkcolor=darkblue,citecolor=darkblue,urlcolor=darkblue]{hyperref}
    \DeclareGraphicsExtensions{.pdf,.png,.eps}
\else
    \usepackage{graphicx}
    \usepackage[hypertex]{hyperref}
    \DeclareGraphicsExtensions{.eps}
\fi

\usepackage{algorithm}
\usepackage[noend]{algpseudocode}
\algnewcommand{\LineComment}[1]{\State \(\triangleright\) #1}   
\renewcommand*\Call[2]{\textproc{#1}(#2)}

\newcommand{\Continue}{{\bf{continue}}}
\usepackage{xfrac} 

\usepackage{fancyhdr} 
\usepackage{subcaption}
\usepackage{enumitem}		
\usepackage{amsmath 
, amssymb  
}
\usepackage{lastpage}
\usepackage{epstopdf}
\usepackage{hyperref}
\usepackage{dblfloatfix}

\graphicspath{{images/}}

\newcommand*{\metaauthori}{Koen De Schepper}
\newcommand*{\metaauthorii}{Olga Albisser}
\newcommand*{\metaauthoriii}{Olivier Tilmans}
\newcommand*{\metaauthoriv}{Bob Briscoe}
\newcommand*{\metashorttitle}{DualQ Coupled AQM}
\newcommand*{\metatitle}{Dual Queue Coupled AQM: \\Deployable Very Low Queuing Delay for All}
\newcommand*{\metano}{TR-BB-2022-001}
\newcommand*{\metakeywords}{Internet, Performance, Queuing Delay, Latency, Scaling, Algorithms, Active Queue Management, AQM, Congestion Control, Congestion Avoidance, Congestion Signalling, Quality of Service, QoS, Incremental Deployment, TCP, Evaluation}
\newcommand*{\metaversion}{01}

\hypersetup{
     pdfauthor = {\metaauthori, \metaauthorii, \metaauthoriii{} and \metaauthoriv},
     pdftitle = {\metashorttitle},
     pdfsubject = {},
     pdfkeywords = {\metakeywords}
}%

\fancypagestyle{first}{%
	\fancyhead[LO,RE]{}%
	\fancyhead[LE,RO]{}%
	\fancyhead[C]{\Large Preprint}%
}%

\begin{document}

\title{\metatitle}

\ifnony{}%
%

\author{
    \metaauthori$^*$\thanks{$^*$Nokia Bell Labs, Belgium, \{koen.de\_schepper$\vert$olivier.tilmans\}@nokia.com}$^\P$
    \metaauthorii$^\dagger$\thanks{$^\dagger$Simula Research, Norway, olga@albisser.org}$^\P$
    \metaauthoriii$^*$
    and \metaauthoriv$^\ddagger$\thanks{$^\ddagger$Independent, UK, research@bobbriscoe.net}\thanks{$^\P$The first two authors contributed equally}
}
\else{}%
\author{
Paper \#101, \pageref{LastPage} pages
}
\fi{}%


\maketitle
\thispagestyle{first}

\begin{abstract}
%
On the Internet, sub-millisecond queueing delay and capacity-seeking have traditionally been considered 
mutually exclusive. We introduce a service that offers both: 
Low Latency Low Loss Scalable throughput (L4S). 
When tested under a wide
range of conditions emulated on a testbed using real residential broadband
equipment, queue delay remained both low (median 100--300\,\(\mu\)s) and consistent
(99th percentile below 2\,ms even under highly dynamic workloads), without
compromising other metrics (zero congestion loss and close to full utilization).
L4S exploits the properties of ‘Scalable’ congestion controls (e.g., DCTCP, TCP Prague). Flows using such congestion control are however very aggressive, which causes a deployment challenge as L4S has to coexist with so-called 'Classic' flows (e.g., Reno, CUBIC). This paper introduces an architectural solution: `Dual Queue Coupled Active Queue Management', which
enables balance between Scalable and Classic flows. 
It counterbalances the more aggressive response of Scalable
flows with more aggressive marking, without having to inspect flow identifiers.
The Dual Queue structure has been implemented 
as a Linux queuing discipline. It acts like a semi-permeable membrane, isolating 
the latency of Scalable and `Classic' traffic, but coupling 
their capacity into a single bandwidth pool. This paper justifies the design and 
implementation choices, and visualizes a representative selection of hundreds of thousands
of experiment runs to test our claims. 


%

\end{abstract}


\section{Introduction}\label{intro}
\subsection{Problem:}%
\label{problem}
With increases in bandwidth, latency is becoming the critical performance factor~\cite{Dukkipati06:RCP_compl_metric, Rajiullah15:LowLatencyUnderstanding}. 
Latency is a multi-faceted problem that has to be tackled on many
different fronts~\cite{Briscoe14b:latency_survey}. 
This paper tackles queuing delay, which is often one of the top contributions to overall latency, alongside propagation delay. 

Although queuing delay is intermittent, as quantified by Hohlfeld \emph{et al}~\cite{Hohlfeld14:QoE_Buffer_Sizing}, it is prevalent enough to dominate experience for two reasons: i) a few higher delay packets often hold back the whole flow of logic~\cite{Briscoe21d:single-delay-metric}; and ii) human perception is dominated by episodes of poorer quality~\cite{Park13:VideoQualityPooling}.

Therefore our goal is very low queuing delay, not just on average but for a high percentile of packets. `Very low' means within single digit milliseconds, for instance 2\,ms at the 99th percentile over a typical Internet path, which is an order of magnitude lower than with PIE~\cite{Pan_PIE_2013} or FQ-CoDel~\cite{Hoeiland18:fq-codel_RFC}.

A major motivation for drastically cutting tail latency is to leave a larger delay budget for propagation, so that interaction can stretch to transcontinental distances. For example, a `responsive feel' in remote presence or remote control needs delay below about 50\,ms~\cite{Carmack13:LatencyMitigationVR}. Foreseeable non-network delays will at best consume 13\,ms of that~\cite{Han17:iccrg-arvr-prob}. Then, reducing P99 queuing from say 30\,ms to 2\,ms (\autoref{fig:ccdf-delay-compare}) would leave 35\,ms rather than 7\,ms for two-way propagation --- increasing reach in fibre from 700\,km to 3500\,km. Not to mention improving the responsiveness of other delay-sensitive applications like online gaming, interactive video and web.

Our further, more ambitious goal is to enable applications that are currently infeasible because they need both high throughput and low delay. These two are generally considered mutually incompatible without significant sacrifice of utilization. This is because high throughput implies capacity-seeking (or capacity adaptation), which has been the main cause of queuing as sources sawtooth or probe for capacity, even under steady-state conditions. Delay variation is particularly acute when the number of flows sharing a link is small (`low stat-mux'), which is the common case when the path bottleneck is in each user's access link. This adds the same order of delay as a typical base round trip time, i.e.\ a peak round trip time (RTT) of about twice the base RTT (see \S\,\ref{intuition}).

To escape this dilemma we need to go beyond an approach like Diffserv Expedited Forwarding (EF~\cite{IETF_RFC3246:EF_PHB}), which offers very low delay but not full utilization --- given it limits the capacity available to the EF class. We also want to go beyond the state-of-the-art in Active Queue Management, such as PIE or FQ-CoDel, which offer high utilization but not very low delay. These AQMs certainly remove excessive queuing, but they still have to buffer the sender's sawtooth variations.

To go beyond these network-only solutions we tackle both sender and network behaviour together. This makes evaluation challenging, because we cannot evaluate one change at a time, and comparisons of all the combinations of sender and network approaches would become too much for one paper. 

Therefore, this paper confines itself to our foundational work on the network part --- the novel `Dual Queue Coupled AQM' (`DualQ') framework. `Foundational' means the network part provides the groundwork --- specifically high-fidelity congestion signals --- on which more dynamic sender behaviours can be built. We only need steady-state traffic over fixed-capacity links to show that the network part works as intended. But we include some dynamic traffic scenarios, albeit only on fixed capacity links, in order to demonstrate the lowest feasible queuing delay under heavy dynamic load. \S\,\ref{scope} fully explains this paper's scope. 

\subsection{Contributions:}%
\label{contributions}

\textbf{Our main contribution} is a new Internet service with an order 
of magnitude lower queuing delay at any percentile, typically without 
sacrificing fixed-capacity utilization.
This is relative to state-of-the-art congestion controls like CUBIC
over state-of-the-art AQMs like PIE or FQ-CoDel.

We call this new service Low Latency, Low Loss, Scalable throughput (L4S). 
It is a general-purpose service for capacity-seeking and other traffic ---
like the Internet's best efforts service, but with very low and consistently 
low queuing delay.

L4S senders can use any one of the family of `Scalable' congestion controls (CCs) (see \S\,\ref{intuition} for the definition of 'Scalable', suffice to say here that DCTCP~\cite{Alizadeh10:DCTCP} is a well-known example). With L4S (as with a DCTCP-enabled data centre), the congestion control at the sender is the primary mechanism that keeps delay low and utilization high, but it also needs shallow unsmoothed Explicit Congestion Notification (ECN~\cite{rfc3168}) in the network (\S\,\ref{ecn}).
For this paper, we use a derivative of DCTCP called Prague~\cite{Briscoe21b:PragueCC-ID}. 

\textbf{Our second contribution} is a solution to the deployability of these Scalable
controls, in coexistence with the traffic already on the public Internet. We use the term `Classic' for this pre-existing traffic sent by Reno-Friendly\footnote{Reno-Friendly is a more precise term for TCP-Friendly.} congestion controls like CUBIC~\cite{cubic} or Reno itself~\cite{IETF_RFC5681:TCP_algorithms}. 

Classic CCs need a decent buffer (holding a typical RTT of data) to absorb their sawtooth window variations without underutilization. To avoid the need for this extra delay, Scalable CCs keep their sawtooth delay variation low by responding much less to each ECN mark relative to a Classic CC. However, `less response' implies `more aggressive', so Scalable flows would outcompete any Classic flows sharing an ECN-capable bottleneck queue.

To solve this `coexistence problem', we propose the DualQ Coupled AQM (`DualQ') that can 
be incrementally added at path bottlenecks. It acts like a semi-permeable
membrane: for delay it isolates L4S traffic from Classic in a separate queue;
but for throughput it couples the queues to appear as a single bandwidth pool
(see \S\,\ref{coupled}). Typical Internet bottlenecks are low stat-mux, so the 
capacity needed (e.g.\ number of flows) in either queue would be highly unpredictable, making it 
hard to allocate appropriate capacity to either. Instead, the coupling enables 
sources to share out the pool between themselves; it couples congestion 
signals across from the Classic to the L4S queue, but to counterbalance 
the more aggressive L4S sources it emits the L4S signals more aggressively.
%

%
\textbf{Our third contribution} is to ensure that the low queuing delay of L4S
packets is preserved during overload from either L4S or Classic traffic, and
neither can harm the other more than they would in a single queue.

We have also tested that either queue can cope
with a reasonable proportion of unresponsive traffic (e.g.\ unresponsive 
streaming, VoIP, DNS), just as classical best efforts can.

\textbf{Our fourth contribution} is a design that needs no reconfiguration once deployed. Two parameters are entirely absent by design, e.g.\ queue smoothing time and dynamic capacity sharing, which are controlled by senders.

\textbf{Our fifth contribution} (in \S\,\ref{eval}) is extensive quantitative evaluation of the above claims; not cherry picking results, but showing all metrics at once: 
i) dramatically reduced delay variability without increasing
other impairments; ii) limited impact on Classic traffic; iii) a good balance between
competing Scalable \& Classic flow rates; and iv) overload handling.

{}


\subsection{Scope:}%
\label{scope}
One paper of reasonable length cannot cover the entirety of L4S in sufficient depth. So, the present paper focuses on the main new component---the DualQ Coupled AQM. This is a framework for coupling AQMs generally. So, to be concrete, we describe the DualPI2 algorithm (pronounced dual-pi-squared), and compare its performance with other state-of-the-art AQMs. A complementary paper provides the theoretical background on coupling Classic \& Scalable AQMs, albeit in just one queue~\cite{DeSchepper16a:PI2}.

Unfortunately, all the following have had to be ruled out of scope in order to keep this paper to a reasonable size:
\begin{description}[nosep]
	\item[Detailed treatment of congestion control algorithms:] Other than a brief overview of the rationale in \S\,\ref{intuition}, and a roadmap of future necessary work in \S\,\ref{cc-reqs}, details of sender CC algorithms are given by reference.
	\item[Per-flow queue (FQ) AQMs with L4S support:] L4S support has been added to the Linux implementation of FQ-CoDel.\footnote{\label{fn:L4S-FQ-CoDel}``fq\_codel: generalise ce\_threshold marking for subset of traffic'', \url{https://git.kernel.org/pub/scm/linux/kernel/git/netdev/net-next.git/commit/?id=dfcb63ce1de6b10b}}{} However, the space needed to explain fine-tuning of the design and to compare with both classical FQ-CoDel and with a DualQ warrants a dedicated paper. 
	Nonetheless, the pros and cons of a DualQ relative to an FQ solution are in scope (\S\,\ref{per-flow}).
	\item[More demanding link types:] The DualQ has been implemented for Data Over Cable (DOCSIS)~\cite{White19a:LowLatencyDOCSIS_Overview} and WiFi links, and simulated for 5G~\iftr\else\cite{Willars21:L4S_5G_etal}\fi{}. However, this paper limits discussion and evaluation to dedicated (e.g.\ DSL, switched Ethernet) rather than shared media link types.
	\item[Coexistence in pre-existing non-L4S queues:] \S\,\ref{Deployment} briefly discusses wider deployment considerations, including 
	how a Scalable control should fall back to Reno-Friendly if it encounters 
	a non-L4S bottleneck and other deployment scenarios
	such as coexistence between Scalable and Classic TCP in heterogeneous or
	interconnected data centres. A technical report is available on detection of whether an L4S flow is bottlenecked at a Classic ECN AQM~\cite{Briscoe19d:ecn-fallback}, and full results with and without detection are available\footnote{\url{https://l4s.net/ecn-fbk/}}.
\end{description}

%


\begin{figure*}[b]
	\vspace{-0.8cm}   
	\centering 
	\begin{subfigure}[b]{\textwidth}
		\centering
		\includegraphics[width=\textwidth,trim={0 8.6cm 3.4cm 0},clip]{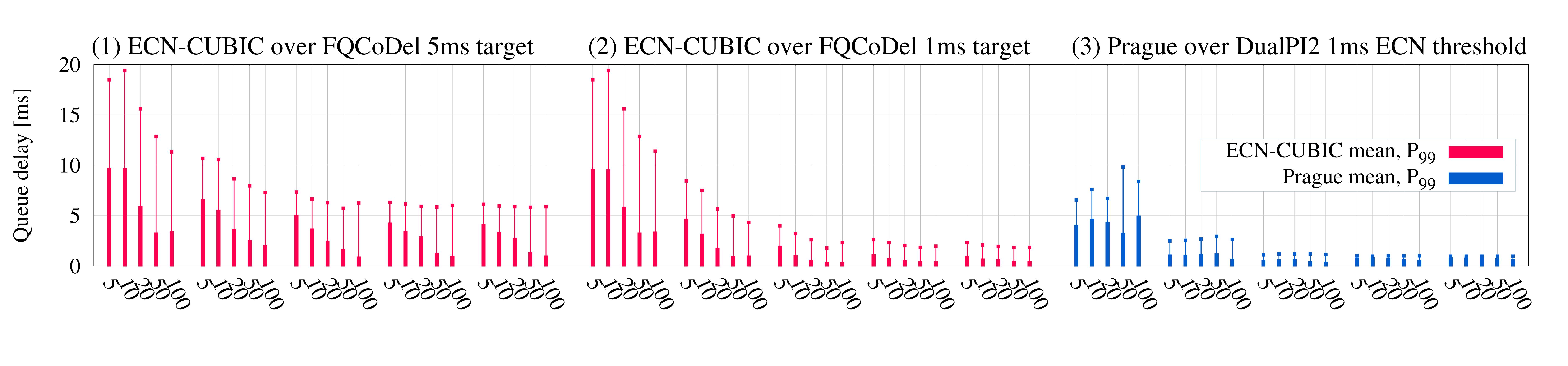}
	\end{subfigure}%
	
	\begin{subfigure}[b]{\textwidth}
		\centering
		\includegraphics[width=\textwidth,trim={1.3cm 3cm 3.4cm 4.3cm},clip]{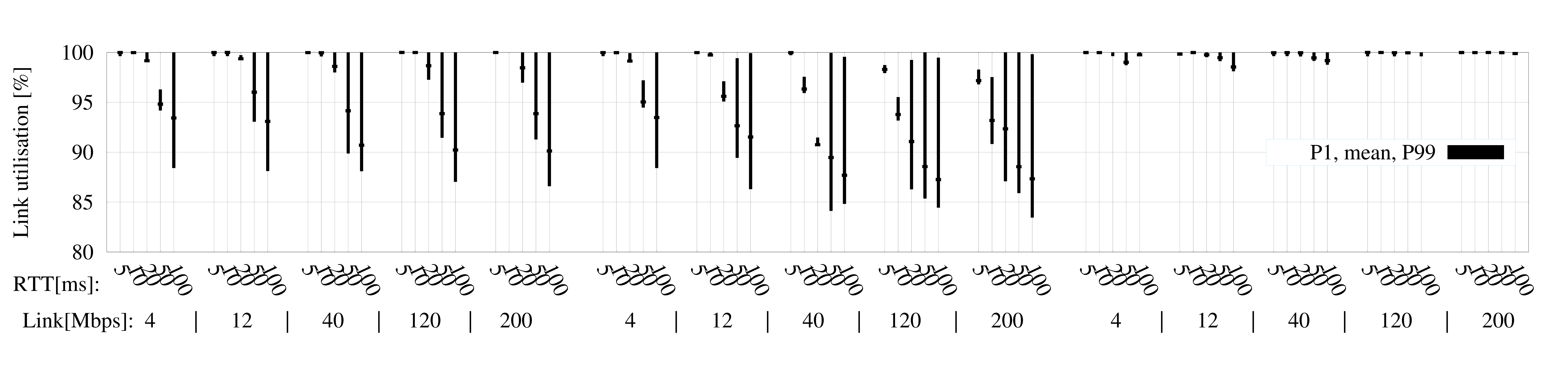}
	\end{subfigure}%
    \addtocounter{figure}{1}
	\caption{With a long-running Classic (ECN-CUBIC) flow, reducing the queue target of the FQ-CoDel AQM from 5\,ms to 1\,ms (left \& middle) makes utilization suffer badly for all but the lowest RTTs. Whereas a Scalable (Prague) flow with small sawteeth can keep to a 1\,ms ECN threshold without compromising utilization (right).}
	\label{fig:efqcodel}
\end{figure*}%

\section{Rationale}\label{rationale}
\subsection{Why a Scalable Congestion Control?}\label{intuition}

A congestion control (CC) is defined as Scalable if, in steady state, the average time from one 
congestion signal to the next (the recovery time)
does not grow as the flow rate scales up, all other factors being
equal~\cite{Briscoe15f:ecn-l4s-id_ID}. Here, `congestion signal' means either loss or ECN, and response to delay is not considered (but not disallowed).

Standard Reno CC~\cite{IETF_RFC5681:TCP_algorithms} has become very slow to recover from any disturbance, because its recovery time (or sawtooth duration) grows linearly with flow rate~\cite{IETF_RFC3649:HSTCP}. CUBIC~\cite{cubic} scales better, but it is still not fully scalable. 
For example, for every 8-fold rate increase, CUBIC's recovery time doubles; e.g.\ from 100 to 800\,Mb/s, recovery time extends from 250 to 500 round trips (for base RTT=20\,ms).

We now derive the condition for a CC to be scalable, at least for those with a response to congestion of the form \(W\propto 1/p^B\). 
Here the steady-state window, \(W\), responds to the probability \(p\) that a packet carries a congestion signal, and \(B\) is a characteristic constant of the algorithm (e.g.\ \(B=\sfrac{1}{2}\) for Reno or \(\sfrac{3}{4}\) for CUBIC).

In steady state, the number of signals per round, \(v\), is the product
of the segments per round \(W\) and the probability \(p\) that a segment carries a
signal, i.e.\ \(v=pW\). Or, substituting for
\(p\) from the response function:
\vspace{-0.2cm}
\begin{align*}
v \propto W^{(1\ -\ \sfrac{1}{B})}.
\end{align*}
The recovery time is the inverse of \(v\). So, by the earlier definition, \(v\) must not reduce as \(W\) increases. Therefore, \((1-\sfrac{1}{B})\ge0\). So, \(B\ge1\) defines a control as Scalable.

For DCTCP~\cite{Alizadeh10:DCTCP} or Prague~\cite{Briscoe21b:PragueCC-ID}, \(B\ge1\), and with probabilistic marking \(B=1\) (see
\S\,\ref{coupled}), so they are Scalable.  The algorithms are designed to induce a high average signalling rate (low recovery time) of two ECN marks per round\footnote{Per `virtual RTT' when Prague is in RTT-independence mode (see \cite{Briscoe21b:PragueCC-ID}).}, and that remains invariant whatever the rate.{}

\begin{figure}
	\centering
	\includegraphics[width=\linewidth]{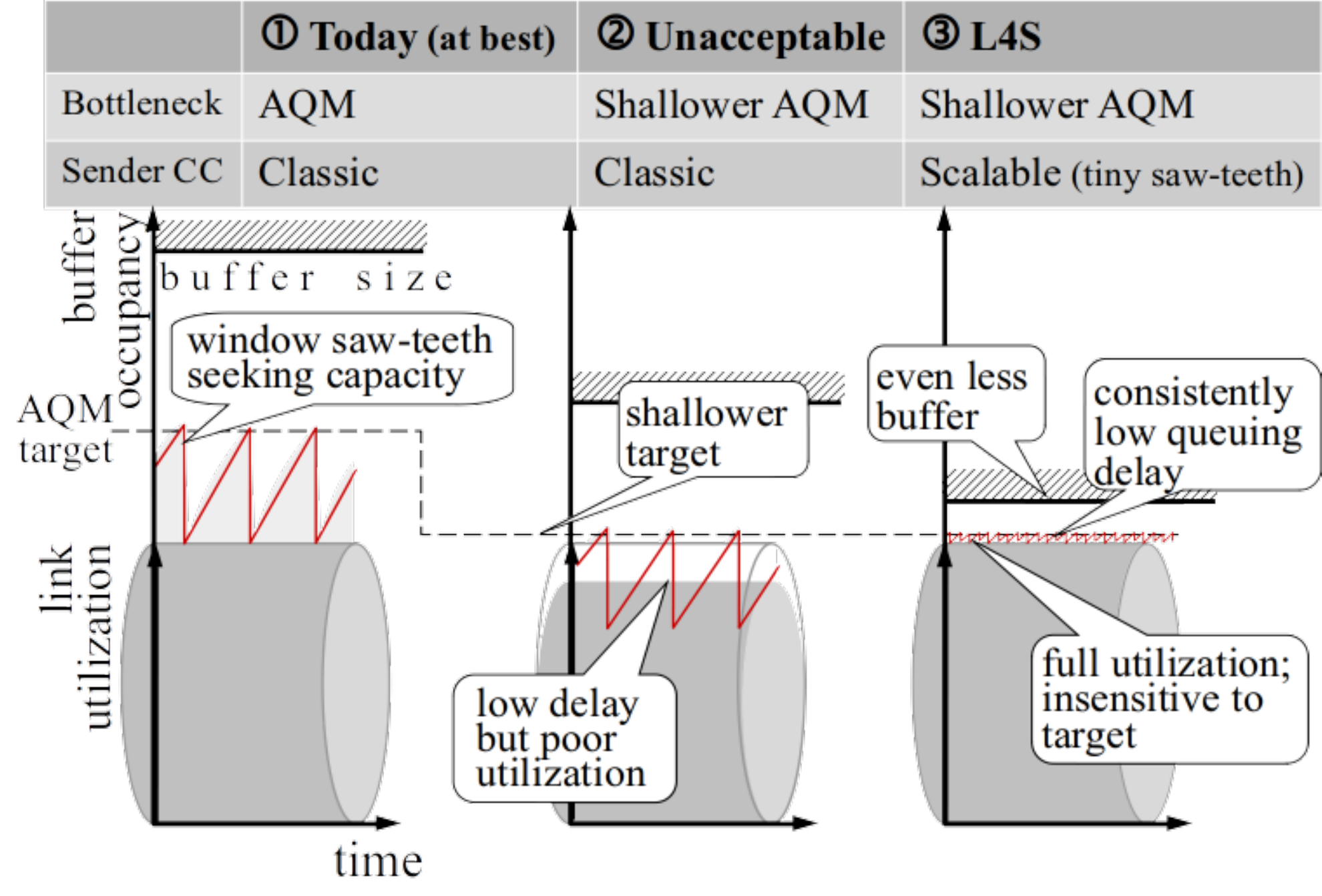}
    \addtocounter{figure}{-2}
	\vspace{-0.1in}
	\caption{Scalable Congestion Control: Intuition}
	\addtocounter{figure}{1}
	\label{fig:dctcp-intuition}
	\vspace{-0.2in}
\end{figure}

To allow the window to scale, it is important to prevent the recovery time from growing so that:
\begin{enumerate}[nosep]
	\item control does not slacken and it does not become more sensitive to noise from transmission losses or transient queuing.
	\item sawtooth amplitude can be small, and stay small.
\end{enumerate}
The schematic in \autoref{fig:dctcp-intuition} gives the intuition for this latter point. With a large amplitude Classic sawtooth, either queue variability is high (1) or under-utilization is excessive (2). But, with low amplitude sawteeth, both can be minimal (3).

The three columns of results in \autoref{fig:efqcodel} reinforce these three points empirically; with CUBIC's large Classic sawteeth ((1) \& (2)), you cannot have both low delay and low under-utilization. But with Prague's tiny sawteeth (3) you can.

The DCTCP analysis~\cite{Alizadeh11:DCTCP_Analysis} calculates that full utilization in steady-state requires the ECN
threshold to be set to \(\ge17\%\) of the RTT. But it also says that utilization is fairly insensitive to this setting. We can confirm that a threshold more than an order of magnitude shallower can be used without noticeable under-utilization.\footnote{Indeed, because the sawteeth are so small, they can either sit just within the buffer, as shown, or just below full utilization, by using a virtual queue, as in HULL~\iftr\else\cite{Alizadeh12:HULL_etal}\fi{}.}

The dynamic behaviours of DCTCP \& Prague are not scalable, mainly
because they still use an additive increase of one segment per RTT, like Reno. However, the dynamic behaviour can be continuously improved after initial deployment (see \S\,\ref{cc-reqs}), whereas the steady-state behaviour is hard to change, once established.

%
For all experiments in this paper, 
TCP Prague~\cite{Briscoe19a:TCP_Prague_Linux} is used in its default configuration\footnote{for source code
of the exact version used, see
\url{https://github.com/L4STeam/linux/blob/66331636f0dd4930/net/ipv4/tcp_prague.c}}. Prague was developed as the reference CC that implements the `Prague L4S Requirements'~\cite{Briscoe15f:ecn-l4s-id_ID}.
It is based on DCTCP, but with the improvements and modifications 
for use over the Internet specified in \cite{Briscoe21b:PragueCC-ID} that have been implemented over TCP and QUIC. 
Other scalable controls have been developed, e.g.\ the ECN-capable part of BBRv2~\cite{Cardwell21:BBRv2_ID} over TCP or QUIC and a scalable variant of SCReAM~\iftr\else\cite{Willars21:L4S_5G_etal}\fi{} for real-time media over RTP.
\bob{Add QUIC Prague when a ref is available.}

\subsection{L4S Explicit Congestion Notification (ECN)}\label{ecn} 

L4S uses the same protocol fields as standard ECN~\cite{rfc3168}, but defines new semantics for an ECN mark. This breaks away
from its previous equivalence to loss, as recently allowed by the IETF~\cite{Black18:ecn-expts}, so that L4S can evade the compromises that are inherent in using
drop as a congestion signal. This is because drop is also an impairment, so it
cannot be signalled too frequently or too immediately. Specifically, L4S
exploits ECN to reduce delay in three respects:
\begin{enumerate}[nosep]
	\item L4S ECN allows more frequent signalling, which would be untenable as loss,
	particularly during high load. This facilitates the smaller sawteeth of scalable
	controllers, which reduce delay as already explained.

	\item When a queue starts to grow, a drop-based AQM holds back from
	introducing loss in case queue growth turns out to be transient. In contrast, an L4S AQM can emit ECN
	immediately, because it is not also an impairment:
    \begin{itemize}[nosep]
		\item With drop, an AQM has to hold back from drop for about 1 RTT. But it
		does not know each flow's RTT, so it has to hold back for a worst-case
		(inter-continental) RTT, to avoid causing instability to worst-case RTT flows.
		
		\item With ECN, the AQM can signal immediately, and the sender can smooth the
		signals---it knows its own RTT, which it can use as the appropriate smoothing time~\cite{Alizadeh11:DCTCP_Analysis} or it can choose to respond without
		smoothing when appropriate, e.g.\ at flow start.
    \end{itemize}

	\item ECN also offers the obvious latency benefit of near-zero
	congestion loss, which primarily benefits short flows, as shown by Salim \& Ahmed~\cite{IETF_RFC2884:ECN_eval}. This removes retransmission and time-out
	delays and the head-of-line blocking that a loss can cause when a transport with ordered delivery (like TCP)
	carries a multiplex of streams.
\end{enumerate}
Because L4S requires Scalable traffic to be ECN-capable, it overloads the 
spare ECN codepoint as the identifier of Scalable packets, for classification into the L4S queue (see \S\,\ref{solution}).

\subsection{Why Is Per-Flow Queuing Not Enough?}\label{per-flow}

Per-flow queuing (as in FQ-CoDel~\cite{Hoeiland18:fq-codel_RFC}) is intended 
to isolate a latency-sensitive flow from the delays induced by others. 
However, FQ alone does not protect a
latency-sensitive flow from the saw-toothing queue that a Classic flow 
still inflicts upon \emph{itself}.%
\footnote{It might seem preferable to release data into a dedicated network
queue, then: a) it would be ready to go as soon as there was capacity; and b)
otherwise the sender would have to hold back the data instead, causing the
same delay, just in a different place. However, modern applications, e.g.\
HTTP/2~\cite{Belshe15:http_2} or interactive video, need to maintain any
self-induced send-queue locally so, at the last moment, they can decide
what to send next dependent on the very latest user behaviour and feedback. 
They cannot alter data already in flight.} This is important for the growing trend of
rate-adaptive interactive video-based apps that are both extremely latency-sensitive 
and capacity-hungry, e.g.\ interactive or conversational video, remote presence.

Support for scalable congestion controls has been added to FQ-CoDel to address the
self-inflicted delay problem.\textsuperscript{\ref{fn:L4S-FQ-CoDel}} It adds shallow-threshold ECN marking for L4S ECN packets while still directly enforcing throughput equality between Scalable and Classic flows (rather than the DualQ's indirect approach).

However, some network operators still consider that some or all of the following are uncomfortable compromises inherent in per-application-flow queuing (FQ):
\begin{description}[nosep]
	\item[Privacy:] Layer-3 VPNs hide transport layer headers for privacy, but FQ needs to inspect flow IDs to isolate flows. So FQ does not allow both flow-privacy and low delay.
	\item[Neutrality:] FQ does not know i) whether a flow using more, or less, than an equal
	share of a user's own capacity is intentional, or even mission-critical; ii)
	whether short-term flow rate variations are deliberate, e.g.\ a more complex
	video scene; iii) whether a real time congestion control is deliberately
	adapting slowly to changing numbers of competing flows, in case they are
	transient.
	\item[Simplicity:] FQ requires more complex classification, queuing and scheduling 
	structures.
\end{description}
The DualQ was developed for those operators that want to offer low 
delay service without being obliged to face these compromises. With a DualQ, an operator can still choose to add a flow policer to enforce equal flow-rates, but as an
independent policy choice, not as a non-optional side-effect of reducing delay.



\begin{figure*}[b]
	\vspace{-0.5cm}
	\centering
	
	\includegraphics[width=\linewidth,trim={0cm 3cm 1cm 6cm},clip]{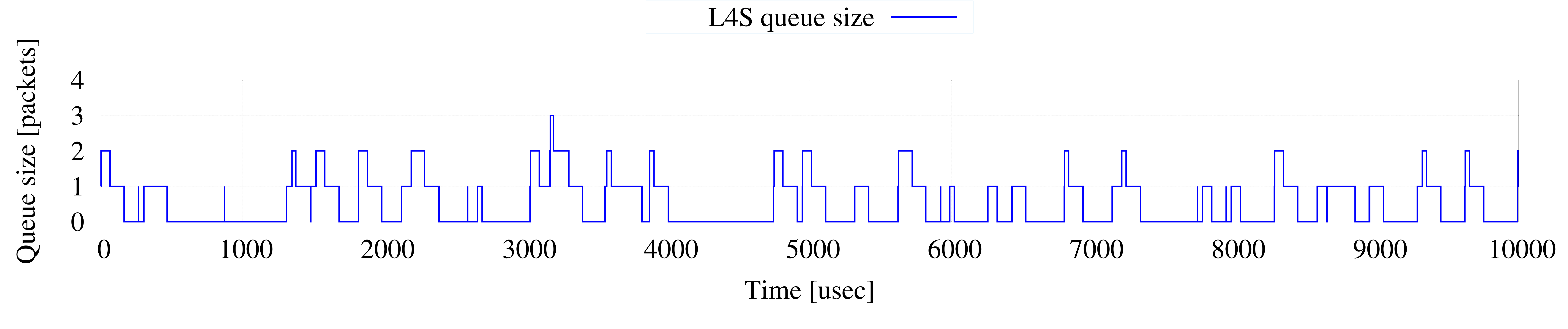}
	\addtocounter{figure}{1}
	\vspace{-0.4cm}
	\caption{L4S queue size over time with 1 long Prague flow and 1 long CUBIC flow in each queue into a 120\,Mb/s link with 10\,ms base RTT. Due to the coupling, the L4S flow leaves the right amount of scheduling opportunities for the Classic queue.}
	\label{fig:qs_timeline}
\end{figure*}%

\section{DualQ Coupled AQM: Solution Design}\label{solution}

\def\RTT{T} 

The solution will be explained in two passes. The first pass introduces the
overall structure (\S\,\ref{solution_structure}). Then details of each aspect
are given in a second pass.

\subsection{Solution Structure}\label{solution_structure}

\paragraph{Latency Isolation} L4S and Classic traffic have opposing delay requirements. The first design
goal of L4S traffic is very low queuing delay. In contrast Classic congestion
controllers (CCs) need a significant queue to avoid under-utilization (and other impairments), in the 
common low stat-mux bottleneck case. One 
queue cannot satisfy these opposing goals, so we use two separate buffers.

Packets are classified between the two queues based on the 2-bit ECN field in
the IP header. Classic sources set the codepoints `ECT(0)' or `Not-ECT'
depending on whether they do or do not support standard (`Classic') ECN~\cite{rfc3168}. L4S
sources ensure their packets are classified into the L4S queue by setting
`ECT(1)', which is an experimental ECN codepoint being redefined for L4S (see
\S\,\ref{codepoints}).

\paragraph{Coexistence} An L4S CC such as Prague achieves low latency, low loss and
low rate variations by driving the network to give it frequent ECN marks. A
Classic CC (Reno, CUBIC, etc.) would starve itself if
confronted with such frequent signals.

So the second design goal is coexistence between Classic and L4S congestion
controllers, meaning rough balance between their
steady-state packet rates.
This problem has already been solved in the single-queue coupled AQM~\cite{DeSchepper16a:PI2} by inflating the
signal intensity applied to L4S traffic to compensate for its reduced response to each
signal. In the DualQ case, the congestion signals that the Classic AQM applies to Classic traffic due to its own queue are also coupled across to be applied to L4S traffic, but with increased intensity (see \S\,\ref{coupled}).

\paragraph{Scheduling}\label{scheduling} Splitting the traffic into two queues raises the question of how often to schedule each queue.
We do not want to schedule based on the number of flows in each, because we want to avoid flow identification (see \S\,\ref{per-flow}). Instead, for the most part, the scheduler is arranged not to interfere, so that sources can `schedule' themselves---just as they would in a FIFO, where the level of congestion signalling rises until each flow draws back just enough to make space for all the other flows. In the DualQ, the coupling of congestion signals from the Classic to the L4S queue makes L4S flows draw back to leave just enough space for the Classic flows.

\autoref{fig:qs_timeline} illustrates this in the example case of one Classic and one L4S flow. It shows only the L queue, to illustrate that the L flow is leaving it empty about half the time.

Bearing in mind that Classic flows tend to build a queue, using a work-conserving strict priority scheduler in favour of the L queue will then be sufficient (superficially, at least). It will serve L packets whenever present (half the time in the example), otherwise it will drain the C queue.

The priority scheduler ensures any queue of C packets yields to L packets. While coupled congestion signals from the C to the L queue make L traffic yield to C. These two opposing mechanisms counterbalance each other, nullifying any bandwidth advantage of priority scheduling.
Nonetheless, to avoid short-term starvation of Classic traffic, priority has to be conditional, not strict, as will be explained in the second pass through the design (\S\,\ref{dualq}). 

\begin{figure}
	\centering
	\includegraphics[width=\linewidth]{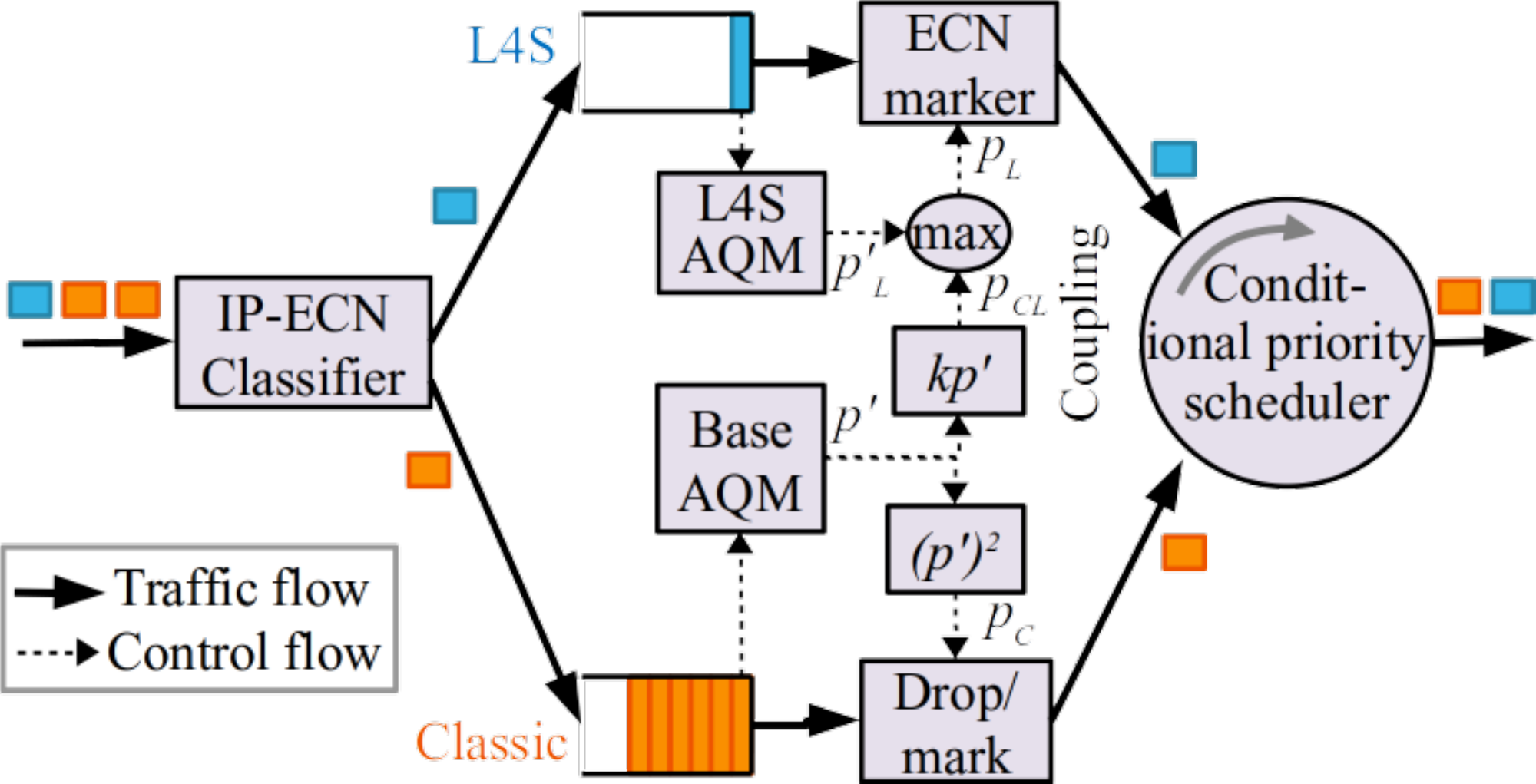}
	\addtocounter{figure}{-2}
	\caption{Dual Queue Coupled AQM: Structure}
	\addtocounter{figure}{1}
	\label{fig:mixcpl3}
	\vspace{-0.5cm}
\end{figure}
\paragraph{The whole picture} The schematic in \autoref{fig:mixcpl3} shows the whole `Dual Queue Coupled AQM', with
the classifier and scheduler as the first and last stages. The detail in between is covered next, but it can be seen that each
queue has its own native AQM that works whether or not the
other queue is empty, while the coupled signals flow across from the Classic AQM to the L4S marker.

\subsection{Coupled AQM for Window Balance}\label{coupled}

To derive the strength of the coupling between
Classic (C) and L4S (L) congestion signals, we take the equations for the
steady-state packet rate, \(r\), of two sender CCs as functions of
congestion signalling probability \(p\). Then, equating the two packet rates
will give the necessary strength of one congestion signal relative to the other.

The packet rate of most CCs also depends on the
end-to-end RTT (\(R=R_b+Q\)), where \(R_b\) is the base RTT and \(Q\) is the
queue delay.  The `fairness' of CCs has generally always been judged under
equivalent conditions, meaning the same RTT, packet size, etc. But with the 
DualQ, a C and an L flow with the same base RTT will have different e2e RTTs, 
because the steady-state queue delay (\(Q\)) is considerably smaller for L flows. Therefore,
before deriving a coupling formula, we need to briefly digress into the RTT-dependence
of L4S and Classic congestion controls.

\paragraph{RTT-dependence} If the rates of two CCs depend inversely on their RTT then, taking
example extreme base RTTs of 4\,ms and 100\,ms, the worst-case rate
ratio will be \((100+Q)/(4+Q)\). If we plug in values of \(Q\) as technology evolves from tail drop buffers (c.100\,ms); through 
state-of-art AQMs (c.10\,ms); to the L4S AQM (\(\sim\)500\,\(\mu\)s), this
worst-case ratio explodes: 2, 8 then 22. Thus, `RTT-unfairness' evolves from a non-problem into
a potential starvation problem. 

But blame does
not lie with lower queuing delay, which has only unmasked the problem. Blame lies with window-based CCs, and that is
where the problem should be solved.

There is no reason why congestion control has to be inversely proportional to
RTT.\footnote{Although it can still depend on RTT outside steady-state.} 
Admittedly, the RTT-dependence of existing Classic traffic cannot be
`undeployed'. Nonetheless, the evaluations in \S\,\ref{expts-diff-RTTs} show that it is possible and sufficient to address
RTT-dependence solely in L4S senders, as also required by \cite{Briscoe15f:ecn-l4s-id_ID}. This addresses the problem for L4S vs.\ L4S
flows, and does not significantly worsen the problem for L4S vs.\ Classic
compared to Classic vs.\ Classic.

\paragraph{Coupling strength} We take the approach of equalizing the rates of L
and C flows at one typical (`reference') base RTT common to both L and C flows
(\(R^*_b\)). As just discussed, flows with significantly different RTT from the
reference should not have significantly different rate: for Classic flows this
is due to the cushioning effect of their queue; and for L4S flows it is by
design of their RTT-independent CC algorithm.%
\footnote{Our worst-case result was the `A5:B100' scenario in
\autoref{fig:mrtt2_link40}, with a rate ratio of 6.3 for 5\,ms Prague vs.\ 
100\,ms Reno over DualPI2. This compares with a ratio of 5.5 for 5\,ms CUBIC vs.\
100\,ms Reno over PIE (a harm ratio of 1.14). We believe this is within our
``not significantly worse'' goal.}

For C and L flows we use the Reno and Prague equations in \autoref{eqn:reno1} and
\autoref{eqn:prague1}, where \(p_{CL}\) is the coupled signal as defined in \autoref{fig:mixcpl3}. We use Reno because it is the worst case (weakest) and it
is widely used, mostly as CUBIC in Reno-compatibility mode (which we call
CReno).{} Coexistence concerns steady-state conditions, so we can
use the simplified steady-state Reno equation from \cite{mathis97}. Prague has
similar steady-state behaviour to DCTCP, but we do not use the equation from the
DCTCP paper~\cite{Alizadeh10:DCTCP}, which is only appropriate for step marking.
Instead, we use the equation that is appropriate to our coupled AQM, where
marking is probabilistic, as derived in Appendix A of \cite{DeSchepper16a:PI2}.
Further, we divide by \(f(R_L)\), which is a generic function that represents the 
various ways that have been proposed to reduce Prague's RTT-dependence~\cite{Briscoe21b:PragueCC-ID, Briscoe17a:CC_Tensions_TR}.

{
	\noindent
	\begin{tabular}{ p{4.0cm} p{3.7cm} }
		\vspace{-0.1in}
		\begin{equation}
		\centering
		r_C = \frac{1}{R_C}\sqrt{\frac{3}{2p_C}}
		\label{eqn:reno1}
		\end{equation}
		&
		\vspace{-0.1in}
		\begin{equation}
		\centering
		r_L = \frac{2}{f(R_L)p_{CL}},
		\label{eqn:prague1}
		\end{equation}
	\end{tabular}
	\vspace{-0.in}
}
Then, equating the two rates at \(R_b^*\) results in \autoref{equ:mixcpl2a}:
\begin{align}
p_C &= \frac{3}{8}\left(\left.\frac{f(R_L)}{R_C}\right|_{R_b^*} \genfrac{}{}{0pt}{}{p_{CL}}{}\right)^2
    &:= \left(\frac{p_{CL}}{k}\right)^2.
\label{equ:mixcpl2a}
\end{align}
We group all the constants into coupling factor \(k\) in \autoref{equ:mixcpl2a}.
In Appendix \ref{coupled-aqms} values of \(k=1.96\) and 2.22 are derived,
depending on whether the Classic CC is Reno or CReno. In our implementation we
use \(k=2\), both because it is the round value for the worst-case (Reno) and
because it is an integer power of 2, which makes implementation efficient.

\paragraph{Classic AQM and Coupling}\label{coupling} The coupling is implemented by structuring the Classic AQM in two stages
(\autoref{fig:mixcpl3}). First what we call a `Base AQM' outputs the internal
probability \(p^{\prime}\). Then \(p^{\prime}\) is transformed depending on
which traffic it is applied to. For Classic traffic it is squared,
\(p_C=(p^{\prime})^2\). But for L4S traffic it is applied linearly, \(p_{CL}
= k*p^{\prime}\). Substituting for \(p^{\prime}\) from the latter into the
former proves that the coupling between \(p_{CL}\) and \(p_C\) will conform to \autoref{equ:mixcpl2a}, as required.

Diversity of Base AQMs is possible and encouraged. Three have been implemented: a variant of RED called Curvy RED~\cite[Appx.\ B]{Briscoe15e:DualQ-Coupled-AQM_ID}, The DualQ specified for Low Latency DOCSIS~\cite{White19a:LowLatencyDOCSIS_Overview} based on PIE, and Linux DualPI2~\cite{Albisser19a:DualPI2_Linux}. They all control queuing time not queue size, given the rate of each queue varies considerably~\iftr\else\cite{Kwon02:Load_v_Queue_AQM}\fi{}. 
This paper assesses our open source DualPI2 implementation in Linux.

The DualPI2 AQM builds on PI\(^2\)~\cite{DeSchepper16a:PI2}, which also 
couples two AQMs to enable coexistence of different CCs. But, unlike PI\(^2\), 
the coupled AQMs are applied in separate queues, with separate delay targets.
Nonetheless, the coupling between the AQMs makes flows behave
as if they are using a single pool of capacity. 
This builds on the theoretical and experimental proof in \cite{DeSchepper16a:PI2} 
that squaring the output of a PI controller
is a more effective, more principled and simpler way of controlling Reno
(rate proportional to $1/\sqrt{p^{\prime}}$) than PI Enhanced
(PIE~\cite{Pan_PIE_2013}). It has the added advantage that the controller's direct (unsquared) 
output can be used to control a scalable CC like Prague as well (rate proportional to
$1/p^{\prime}$). 
The derivation of the coupling formula is more straightforward in the single-queue case of \cite{DeSchepper16a:PI2}, 
because C and L flows share a common queue delay, so the RTTs cancel out.

\subsection{Dual Queue for Low Latency}\label{dualq}

\paragraph{Native L4S AQM} Often, there will only be traffic in one queue, so each queue needs its own
native AQM. The L4S queue keeps delay low using a shallow marking threshold
(\(T\)), which has already been proven for DCTCP. However, unlike DCTCP, \(T\)
is cast in units of time~\cite{Kwon02:Load_v_Queue_AQM, Bai16:ECN_GPS}, so that
it is invariant with dequeue rate, which can vary considerably.
In case of a low rate link, \(T\) has a floor of two packets. 
On-off marking may~\cite{6681614} or may 
not~\cite[\S5]{Kuzmanovic05:ECN_SYN_ACK} be prone to instability. But to test 
one change at a time we defer investigation of alternatives such as a ramp to future research.

If there is traffic in both queues, an L4S packet can be marked either by the
native L4S AQM or by the coupling from the Classic AQM, whichever outputs higher probability (illustrated as 
the max() function in \autoref{fig:mixcpl3}).
Marking via the coupling generally ensures that L4S traffic stays below its native 
threshold (\autoref{fig:qs_timeline}), only touching it during bursts or if there is insufficient Classic traffic.

Note that Classic AQMs filter out rapid variations in the queue before
it drives dropping or marking, which is necessary to stabilize Classic CCs, but it delays the signals. 
In contrast, as \S\,\ref{ecn} explained, the L4S AQM emits ECN marks without
delay, and L4S sources only smooth the marks if they need to, e.g.\ during their congestion avoidance phase.

\paragraph{Conditional Priority Scheduler} Earlier, an explanation was promised for why the scheduler's priority has to be conditional. We found that strict priority worked fine in steady-state, but sometimes there would be a brief deadlock when a DNS request or the initial packet(s) of a Classic flow arrived at an empty C queue, while L traffic was keeping the L queue busy. This was because, no matter how long a blocked C packet waited, our implementation didn't increase the coupled marking of L packets, because it only measured a packet's delay when it was dequeued, not while it was waiting. The C packets were only ultimately released when there happened to be a gap in the L traffic.


Conditional priority, such as a 
weighted round robin (WRR) scheduler with a high weight 
in favour of the L queue (e.g.\ \(\sfrac{15}{16}\) or \(\sfrac{9}{10}\)) resolves this deadlock. The actual weight is immaterial, because capacity shares are determined by the response of L4S senders to the coupled signal (not by the scheduler). The weight for Classic traffic only has to be large enough (e.g.\ \(\sfrac{1}{16}\) or \(\sfrac{1}{10}\)) to release the deadlock.

We also tried a MEDF scheduler~\cite{Menth03:MEDF}, which we call a Time-Shifted FIFO. 
It selects the packet with the earliest arrival timestamp, after subtracting a 
constant time-shift to favour L4S packets. It performs nearly as well as WRR 
despite its simplicity. However, it allows bursts of delay to 
leak from the C to the L queue, so it is not used further in this paper. 

%

\subsection{Overload Handling}
\label{overload}
Having introduced a priority scheduler, during overload we must at least 
ensure that it gives unresponsive traffic no more power to harm other 
traffic than a single queue would. 
We
prefer to leave flow policing as a policy choice (see \S\,\ref{per-flow}).
So, the coupled AQM allows unresponsive traffic below the link rate to just subtract from the overall
capacity, whether it classifies itself as L4S or
Classic. Then it still allows any responsive
flows to share the remaining capacity ---
as they would in a single queue with the same capacity
subtracted.

To handle excessive unresponsive traffic, the Base AQM is actually driven by
whichever queue is greater\footnote{Even if the L4S queue is the greater, e.g.\ no C traffic, when not overloaded, it stays well below the Classic target. So it drives \(p^\prime\) to zero.}, and when its output probability exceeds a threshold it applies
drop to ECN-capable packets in either queue with the same probability as Classic
(i.e.\ the coupling becomes bidirectional and equal). By default, this overload
threshold is set where the L4S AQM saturates at 100\%
marking. By equation (\ref{equ:mixcpl2a})
this occurs once Classic drop reaches \((100\%/k)^2\), which is \(25\%\) if \(k=2\). 

When any L4S
source detects a drop, it is required to react as a classic flow would~\cite{Briscoe15f:ecn-l4s-id_ID}, so balance
between flow rates is preserved.
The L4S AQM also continues to ECN-mark 
so that, under unresponsive Classic load, any responsive L4S traffic can maintain the very low queuing
delay of the L4S service. These claims are verified in \S\,\ref{expts-overload}.

\subsection{Implementation}
\label{implementation}

{}
\begin{algorithm}
\begin{algorithmic}[1]
\small
\While{\Call{lq.len}{} + \Call{cq.len}{} \textgreater 0}
 \If{\Call{scheduler}{} ==  LQ}\label{line:sch}
  \State \Call{lq.dequeue}{pkt}	\Comment{L4S}
  \State $p^{\prime}_L$ = \Call{laqm}{\Call{lq.time}{}}\label{line:laqm}
  \State $p_L$ = \Call{max}{$p^{\prime}_L, p_{CL}$}\label{line:max}
  \If{\(p_L > \Call{rand}{}\)}\label{line:mark_L}
   \State \Call{mark}{pkt}
  \EndIf
 \Else
  \State \Call{cq.dequeue}{pkt}	\Comment{Classic}
  \If{ \(p_C > \Call{rand}{}\) }\label{line:square_p}
   \If{\Call{ecn}{pkt}==0} \Comment{Not ECT}\label{line:notecn_C}
    \State \Call{drop}{pkt} \Comment{Squared drop}
    \State \Continue            \Comment{Redo loop}
   \Else 								  \Comment{ECT(0)}
    \State \Call{mark}{pkt} \Comment{Squared mark}
   \EndIf
  \EndIf
 \EndIf
 \State \Call{return}{pkt} \Comment{return the packet, stop here}
\EndWhile
\end{algorithmic}
\caption{Dequeue for Dual Queue Coupled AQM}
\label{alg:dual-coupled}
\end{algorithm}

\iftr
\else{}%
Algorithm \ref{alg:dual-coupled} summarizes the per packet dequeue implementation %
\fi{}%
of DualPI2 as pseudocode. 
The AQMs are applied at dequeue to minimize signalling delay. For clarity,
overload logic and edge cases are omitted, but they can be found in the  
open-sourced implementation of the DualPI2 Linux qdisc~\cite{Albisser19a:DualPI2_Linux}, or as pseudocode in \cite[Appx.\ A.2]{Briscoe15e:DualQ-Coupled-AQM_ID}. 
The function
\textsc{len()} returns the the queue in bytes, while \textsc{time()} returns
the duration since a packet was time-stamped (its sojourn time).

\iftr
\else{}%
On enqueue (not shown), packets are time-stamped and classified based on the least significant bit of the IP-ECN field. 
\fi{}%

On dequeue, line \ref{line:sch} determines which head packet to take. For this
paper we use WRR as already explained, but the pseudocode is generalized 
for any work-conserving scheduler.

If an L4S packet is scheduled, line \ref{line:laqm} runs the native L4S AQM (referred to as \textsc{laqm()}) to
output probability \(p^{\prime}_L\) dependent on the service time of the L queue. This is
a generalization for whatever native L4S AQM is used, but for the present
paper we use a simple [0,1] step function at delay threshold \(T\). Line
\ref{line:mark_L} marks the packet if a random marking decision is drawn
according to the probability \(p_L\), which the previous line has taken as the
max of the outputs of the native L4S AQM (\(p^{\prime}_L\)) and the coupling (\(p_{CL}\)). The latter is maintained by Algorithm \autoref{alg:dual-coupled-updateP}.

If a Classic packet is scheduled, line \ref{line:square_p} decides whether to emit a 
congestion signal with probability \(p_C\), which is also maintained by Algorithm \autoref{alg:dual-coupled-updateP}, as explained in \S\,\ref{coupled}c). Then line \ref{line:notecn_C} checks 
whether the Classic packet is not ECN-capable, in which case it uses drop
as the signal, otherwise it uses ECN.

The smoothed marking and dropping probabilities are kept up to date by Algorithm
\ref{alg:dual-coupled-updateP} which only needs occasional execution~\cite{Hollot02:PI_AQM} every \(T_{\mathrm{update}}\) (default 16\,ms). Line \ref{line:maxq} bases the algorithm on the max of the two queues, which handles overload as explained earlier. 
Then the core PI algorithm updates the internal base signalling probability (\(p^{\prime}\)). The
change in queuing time is multiplied by the proportional gain factor $\beta$. The
integral gain factor $\alpha$ is typically smaller, to restore the delay of any persistent
standing queue to the target. These expressions, which can
be negative, are added to the previous \(p^{\prime}\) every \(T_{\mathrm{update}}\). 
Then the Coupled and Classic signalling probabilities, \(p_{CL}\) and
\(p_C\), are derived from \(p^{\prime}\) (see \S\,\ref{coupled}).
\begin{algorithm}
\begin{algorithmic}[1]
\small
  \State $curq = \Call{max}{\Call{cq.time}{}, \Call{lq.time}{}}$\label{line:maxq}
  \State $p^{\prime} = p^{\prime} + \alpha * (curq - TARGET) + \beta * (curq - prevq)$
  \State$p_{CL} = k * p^{\prime}$
  \State$p_C = (p^{\prime})^2$
  \State $prevq = curq$
\end{algorithmic}
\caption{PI core: Every \(T_{\mathrm{update}}\) \(p\) is updated}
\label{alg:dual-coupled-updateP}
\end{algorithm}

\section{Evaluation}\label{eval}

\subsection{Testbed Setup}\label{tebset}

Experiments were first conducted on a testbed that used realistic DSL equipment, then on a simpler testbed capable of extending over a wider range of BDPs. 
The DSL testbed was assembled using carrier grade equipment in the same
environment as for testing customer solutions, consisting of a classical residential service delivery
network composed of Residential Gateway, xDSL DSLAM (DSL Access Multiplexer),
BNG (Broadband Network Gateway), Service Routers (SR) and application servers. The
Residential Gateway was connected by VDSL to a DSLAM, which was connected
to the BNG through an aggregation network, representing a local ISP or access
wholesaler. Traffic was routed to another network representing a global ISP that
hosted the application servers and offered breakout to the Internet. {} The Linux client and server in either pair A or B, were
always configured with the same TCP variants and applications.

In a production access network, the BNG is usually deliberately arranged as the
downstream bottleneck for each customer (to confine any QoS handling to one point).
Traffic from the client-server pairs was routed from the downstream interface of the BNG through a Linux 
`AQM server', which acted as the BNG's rate bottleneck where we configured
different AQMs to be evaluated. This server also added extra
delay, controlled the experiments, captured the traffic and analysed it. In practice 
it would also be important to deploy an AQM in the home gateway, but in our 
experiments the ACK traffic was below the upstream capacity.

The xDSL line was configured at 48\,Mb/s downstream and 12\,Mb/s upstream. Ethernet links
between network elements consisted of at least 1GigE connections, except 100\,Mb/s between clients and modem. The base RTT between the clients and servers was $7\,\mathrm{ms}$, which was primarily
due to the interleaved Forward Error Correction (FEC) configured for xDSL. Extra base RTT was configured 
using a netem qdisc on the upstream AQM server interface or on the clients for experiments with different base RTTs.

\begin{figure}[H]
	\vspace{-0.15in}
	\centering
	\includegraphics[width=\linewidth]{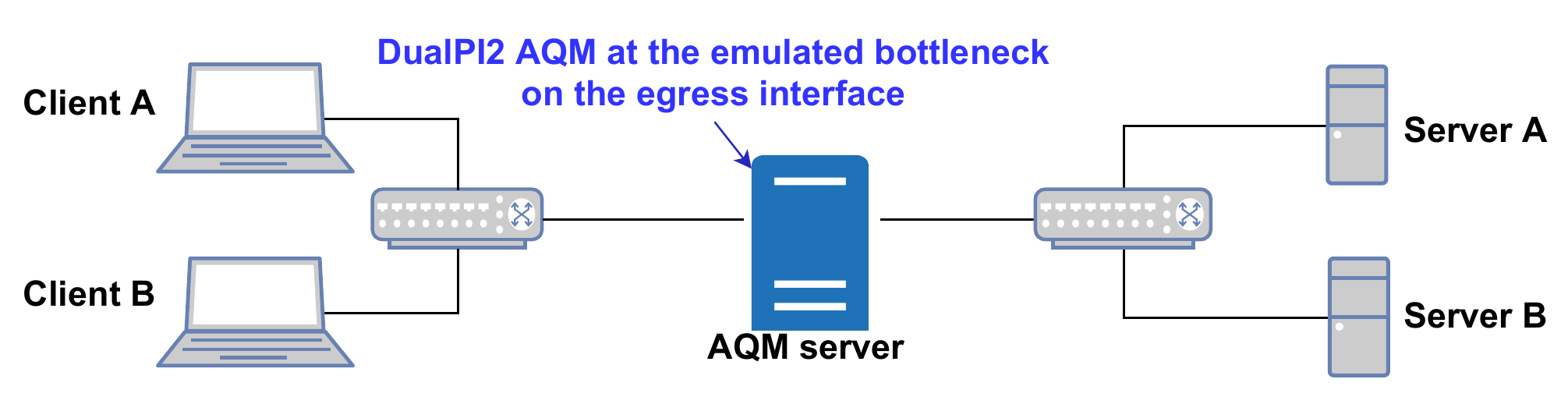}
	\caption{Simpler testbed topology}
	\label{fig:simpler_testbed}
	\vspace{-0.1in}
\end{figure}

All experiments presented in this paper were performed on the simpler testbed, which consisted of 5 nodes (2 clients, 2 servers, an AQM server) and 2 dedicated switches, each connecting the clients and servers to the AQM server with 1GigE connections (shown in \autoref{fig:simpler_testbed}). A consistency check between the two testbeds within the narrower limits of the xDSL testbed showed near-identical results. 
All Linux computers were Ubuntu 18.04/20.04
LTS, running an out-of-tree kernel containing
TCP Prague and DualPI2.\footnote{5.10.31-3cc3851880a1-prague-37 see \url{https://github.com/L4STeam/linux/tree/3cc3851880a1%b8fac49d56ed1441deef2844d405
}} 
We used Prague for the Scalable congestion control and CUBIC for
Classic, both were bundled with the installed kernel version and used with default configuration. For ECN-CUBIC, we enabled TCP ECN negotiation.
We compared DualPI2 with PIE and FQ-CoDel, all configured as in Table~\ref{tab:aqm-pars}. The $\alpha$ and $\beta$ values for PIE are equivalent to those used for DualPI2, but PIE scales the input parameters internally. All AQM parameters were left at their default values.
\begin{table}[h]
\centering
\begin{tabular}{| l | p{6.6cm} |}
\hline
All & Buffer: 40,000\,pkt, ECN enabled \\
\hline
PIE              & Target delay: 15\,ms, Burst: 100\,ms, TUpdate: 16\,ms, $\alpha$: 1/16, $\beta$: 10/16, ECN\_drop: 25\%\\
\hline
FQ-CoDel    & Target delay: 5\,ms, Burst: 100\,ms\\
\hline
DualPI2      & Target delay: 15\,ms, TUpdate: 16\,ms, L4S T: 1\,ms, Classic weight: 10\%, $\alpha$: 0.16, $\beta$: 3.2, k: 2, drop\_on\_overload\\
\hline
\end{tabular}
\caption{Configuration parameters for the different AQMs.}
\label{tab:aqm-pars}
\vspace{-0.2cm}
\end{table}

\subsection{Experimental Approach}\label{exp_appr}

The set of experiments was constructed to evaluate
our main performance goals: queuing delay, utilization and rate balance for most experiments, and flow completion efficiency for experiments with short flows.\footnote{Source data for all presented experiments is available in \url{https://github.com/olgaalb/dualpi2eval}} 
We also show window balance and drop/mark probability as secondary evaluation metrics, mainly to support interpretation of the primary effects. The evaluation metrics are fully specified in Appendix\,\ref{metrics}.

For traffic load we used long-running flows
(\S\S\,\ref{expts-basic-steady-state}~to~\ref{expts-diff-RTTs}) and/or dynamic short
flows (\S\,\ref{expts-dynamic}). We used long flows, not as an example of a 
realistic Internet traffic mix, rather to aid interpretation of various effects, such 
as starvation. Heavy load scenarios predominate in our choice of experiments, again not because they are typical, but because they do occur and they are a challenging case.

We mixed different numbers of flows, different congestion
controls (CCs) and different RTTs. Also, to verify behaviour with unresponsive flows and overload
(\S\,\ref{expts-overload}), we injected unresponsive UDP load; both below and above link capacity, and both ECN and
Not-ECN capable.

As explained in \S\,\ref{scope}, we have reported separately on our experiments with combinations of CCs that an AQM is not intended to support.\footnote{See \url{https://l4s.net/ecn-fbk/results_v2.2/full_heatmap_rrr/} and \cite{Briscoe19d:ecn-fallback}}
Here we choose examples of typical combinations that are supported: Prague with CUBIC on DualPI2\footnote{We do not report results with ECN-CUBIC over DualPI2, which were hardly any different to CUBIC.} and ECN-CUBIC 
with CUBIC on current AQMs (PIE and FQ-CoDel). {}In some cases, we added experiments with Reno instead of CUBIC to check that there was little difference, given CUBIC remains largely in its Reno-friendly 
mode with the BDPs used in these experiments, which were chosen as typical of `data centre to the home' scenarios.

Most of our evaluation is comparative, showing the results for each of the AQM and CC combination, apart from the scenario with unresponsive flows and overload, where only results for DualPI2 are presented\footnote{Comparative overload experiments against the other AQMs are available at \url{https://l4steam.github.io/overload-results/}}.

\subsection{Basic Steady State Experiments}\label{expts-basic-steady-state}
\begin{figure*}[!ht]
    \centering
    \begin{subfigure}[b]{\textwidth}
        \centering
        \includegraphics[width=0.98\textwidth,trim={1.3cm 0 3.4cm 0.3cm},clip]{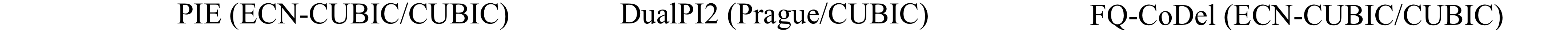}
    \end{subfigure}%
     \vspace{-0.3cm}
    
    \begin{subfigure}[b]{\textwidth}
        \centering
        \includegraphics[width=0.98\textwidth,trim={1.3cm 3cm 3.5cm 0},clip]{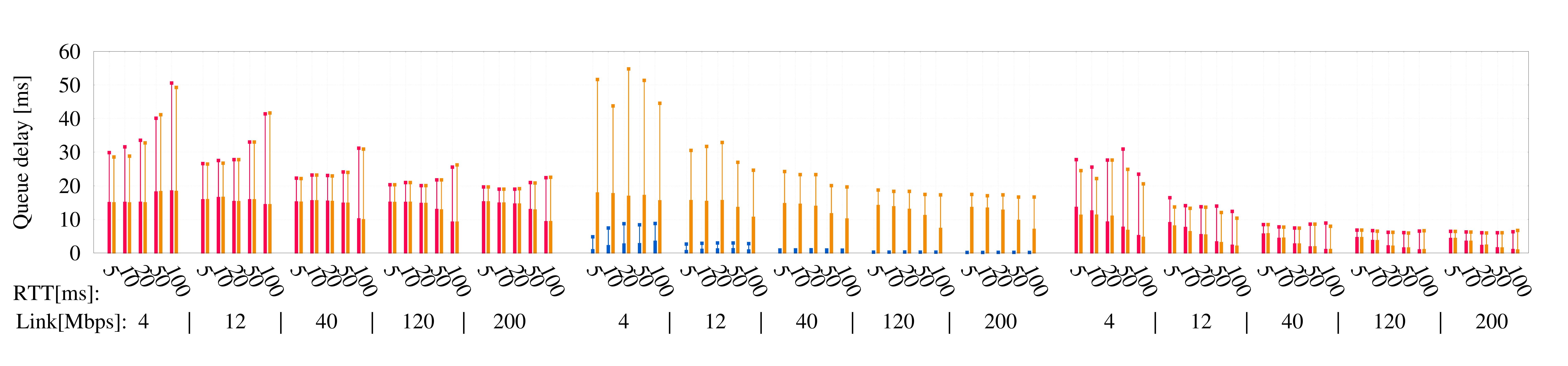}
    \end{subfigure}%

    \begin{subfigure}[b]{\textwidth}
        \centering
        \includegraphics[width=0.98\textwidth,trim={1.3cm 2cm 3.6cm 1.5},clip]{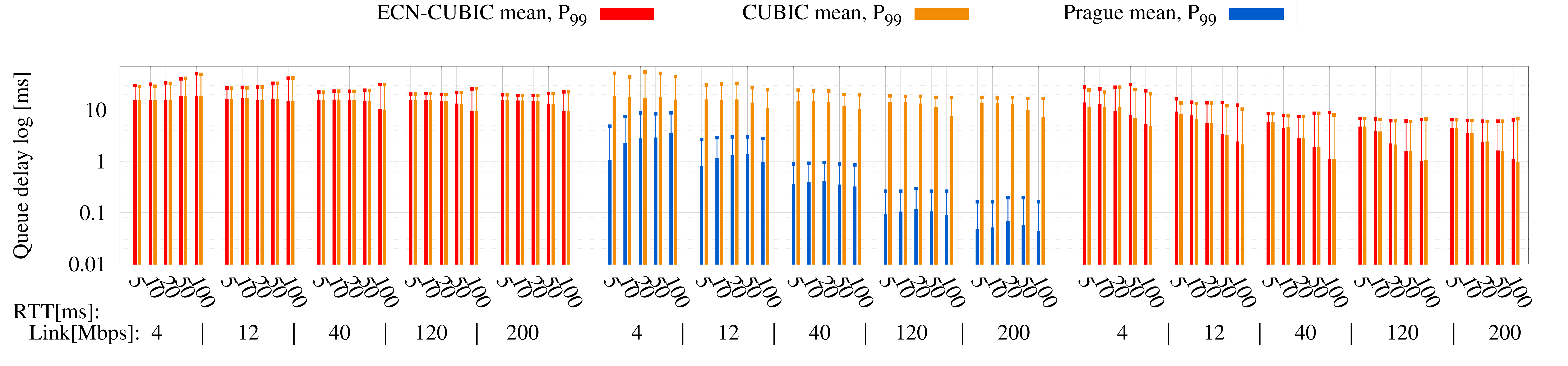}
    \end{subfigure}%

    \begin{subfigure}[b]{\textwidth}
	    \centering
	    \includegraphics[width=0.98\textwidth,trim={1.3cm 0 3.4cm 1cm},clip]{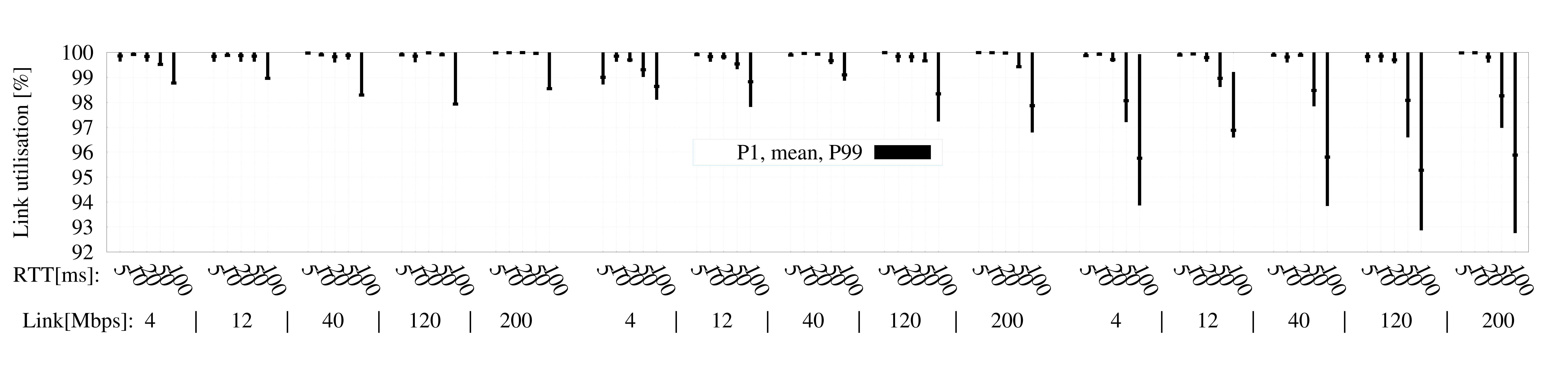}
    \end{subfigure}%

    \begin{subfigure}[b]{\textwidth}
        \centering
        \includegraphics[width=0.98\textwidth,trim={1.3cm 3cm 3.5cm	 2.5cm},clip]{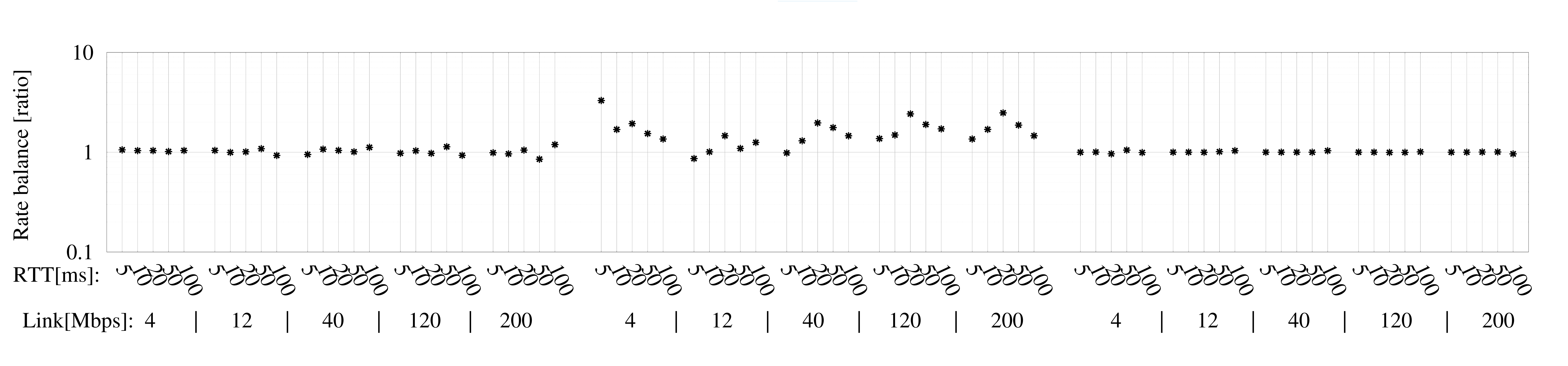}
    \end{subfigure}%

    \begin{subfigure}[b]{\textwidth}
    \centering
        \includegraphics[width=0.98\textwidth,trim={1.3cm 3cm 3.5cm 2.5cm},clip]{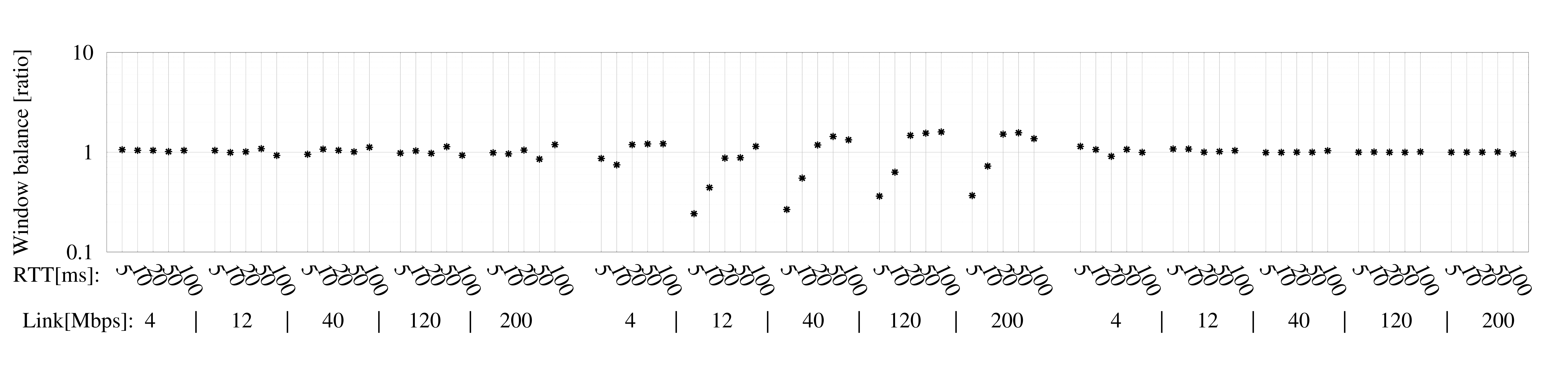}
    \end{subfigure}%

    \begin{subfigure}[b]{\textwidth}
        \centering
        \includegraphics[width=0.98\textwidth,trim={1.3cm 0 3.4cm 0},clip]{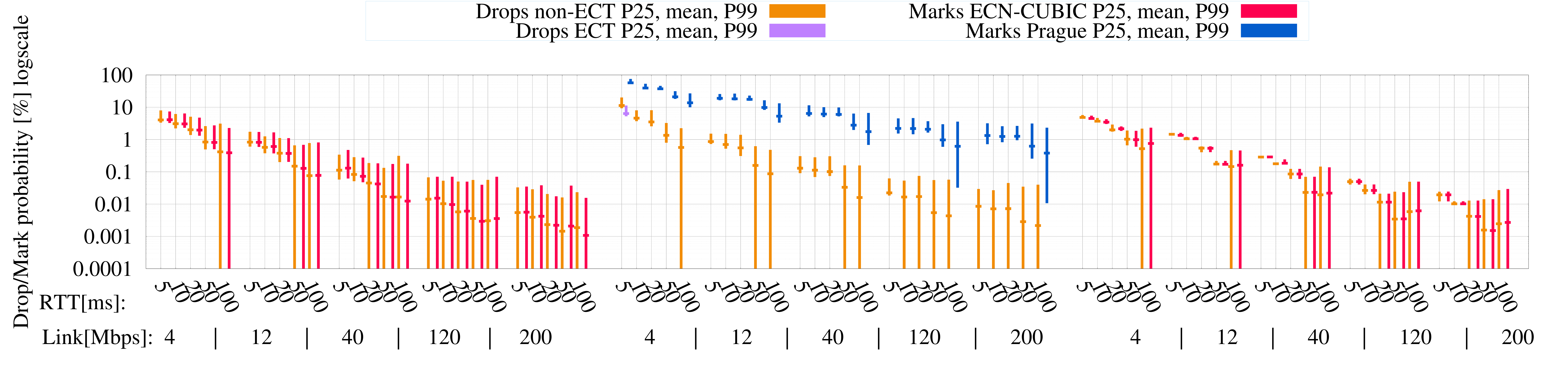}
    \end{subfigure}%

    \caption{Basic Experiments: DualPI2 compared to PIE \& FQ-CoDel; 1 steady-state flow for each CC with equal base RTT}
    \label{fig:1-1_aqm}
\end{figure*}

\paragraph{Experimental setup} In the first set of experiments the competing flows in each experiment had equal base RTTs. For the most basic setup, we used two long running flows; one for each CC and ran experiments with all 
25 combinations of 5 base RTTs (5--100\,ms) and 5 link speeds (4--200\,Mb/s) in each of the three columns; one for each AQM, totalling to the 75 scenarios across \autoref{fig:1-1_aqm}. Each row of the figure reports a different metric: queue delay (linear and log-scale), link utilization, rate \& window ratios and mark/drop probability.

\paragraph {Results} 
The primary goal of the DualQ is very low queue delay for L4S traffic, while ensuring coexistence with Classic traffic. So, compared with Classic AQMs, Classic queueing delay is not expected to be better, but it should be no worse. It is also important to keep a reasonable balance between flow rates despite the different RTTs.

Overall, \autoref{fig:1-1_aqm} shows that indeed DualPI2 enables scalable congestion controls (Prague) to achieve an order of magnitude lower queueing delay than the other AQMs with minimal impact on Classic traffic. Importantly, P99 queue delay is also an order of magnitude lower.

The log plot in the second row of \autoref{fig:1-1_aqm} shows how the DualPI2 \textbf{queue delay} approaches the minimum possible --- the mean at each link rate translates to 1 packet of serialization delay and each 99th percentile translates to 2--3 packets --- consistent with the example in \autoref{fig:qs_timeline}. 

This very low delay is achieved with hardly any harm to the Classic traffic. DualPI2 keeps average Classic delay (CUBIC) mostly at the expected target of 15\,ms. The 99th percentile is somewhat higher than PIE for the lowest link rates, but it is lower in other cases. Although the Classic queuing delay of DualPI2 is not as low as that of FQ-CoDel, DualPI2 does not underutilize the link as much as FQ-CoDel. As was shown in \autoref{fig:efqcodel}, Classic congestion controls create a dilemma between queue delay and under-utilization. So, the target delay of PIE or PI2 could be reduced from the default configuration to match the delay results of FQ-CoDel,{} but only by choosing to sacrifice more utilization, as FQ-CoDel does.

The fourth and fifth rows of \autoref{fig:1-1_aqm} show \textbf{rate} and \textbf{window} balance, which are almost perfect for PIE and FQ-CoDel. This is unsurprising given the two flows are essentially identical (except one supports Classic ECN, which is not expected to alter the flow rate). For FQ-CoDel the results are even flatter due to the round robin scheduling.

The rate balance of flows with the same base RTT is not so perfect over DualPI2 because the greater Classic queue causes greater total RTT and Classic congestion controls are RTT-dependent. As explained in \S\,\ref{coupled}, the proper place to reduce RTT-dependence is in newly deployed L4S congestion controls, and that has indeed been implemented in Prague.

Over the range of scenarios tested, it can be seen that Prague keeps the rate ratio within acceptable bounds; roughly between 0.85 \& 2.5, except for the very smallest BDP\footnote{Here  DualPI2 occasionally tips into overload mode, where the last row of plots shows that some drop is applied to the L4S queue. Prague's RTT-independence algorithm scales down the window so currently it hits the minimum window sooner, causing the higher rate ratio shown. It is planned to reduce the minimum window to compensate for this effect.}. Prague's RTT-independence kicks in below 25\,ms. This can best be seen in the plots of window balance which stay relatively flat as base RTT is reduced from 100\,ms. Then the points drop away below 20\,ms. Equivalently, as RTT reduces, the rate balance plots rise but then fall back towards 1 for RTTs below 20\,ms.


The last plot of \autoref{fig:1-1_aqm} shows how each AQM applies mark and drop probability. It is clearly visible that DualPI2 doubles the signal for Prague and squares it for CUBIC, whereas PIE and FQ-CoDel apply the same signal to both flows. 

We conclude that in steady state DualPI2 can achieve near minimum mean and P99 queue delay without compromising utilization and preserving approximate rate fairness.

\subsection{Multiple long-running flows}\label{multiple-flows}

\begin{figure*}[!ht]
    \centering 
    \begin{subfigure}[b]{\textwidth}
        \centering
        \includegraphics[width=\textwidth,trim={-2cm 0 0 0},clip]{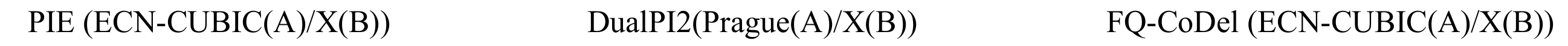}
    \end{subfigure}%
    \vspace{-0.5cm}

    \begin{subfigure}[b]{\textwidth}
        \centering
        \includegraphics[width=\linewidth,trim={0cm 14.1cm 2.8cm 0},clip]{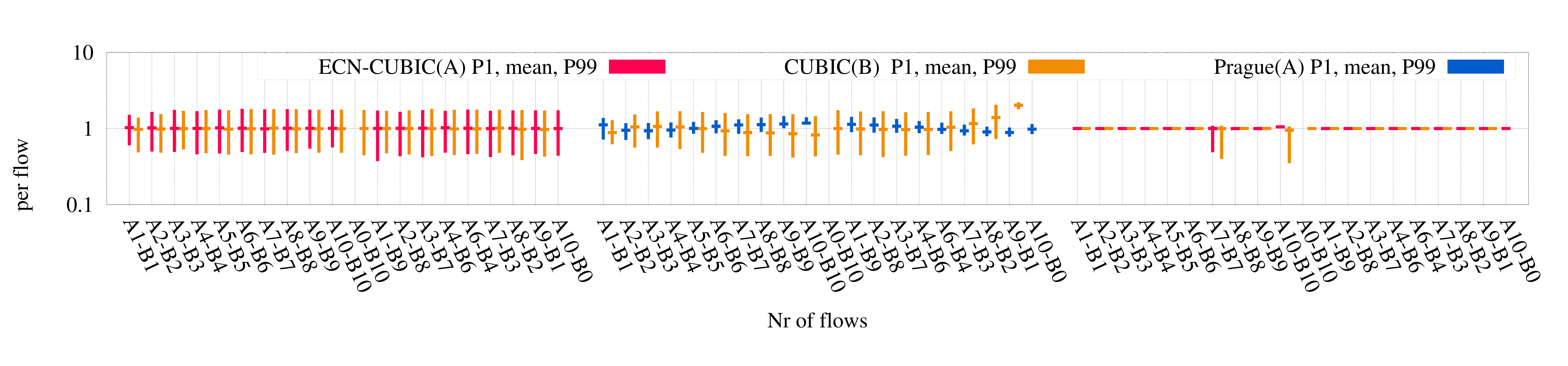}
    \end{subfigure}%

    \begin{subfigure}[b]{\textwidth}
        \centering
        \includegraphics[width=\linewidth,trim={0cm 2cm 2.7cm 4cm},clip]{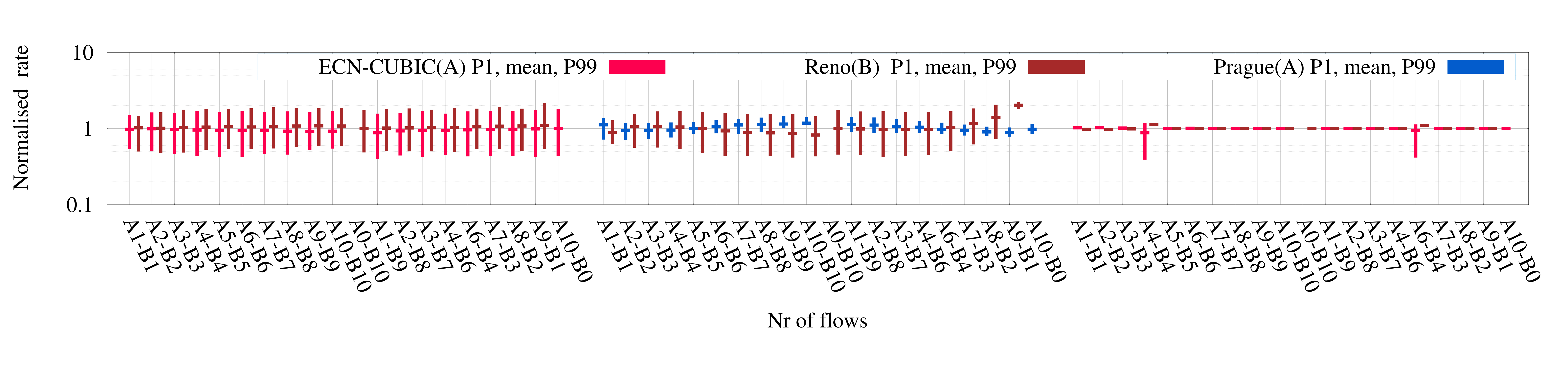}
    \end{subfigure}%

	\vspace{-0.2cm}
    \caption{Set-up as \autoref{fig:1-1_aqm} but with different numbers of flows; Link: 40\,Mb/s; Base RTT: 10\,ms. Queue delay and utilization (not shown) are similar to \autoref{fig:1-1_aqm}}
    \label{fig:l40r10}
\end{figure*}

\paragraph{Experimental setup} As in the basic steady-state experiments, this scenario still 
involves the same two types of flows (ECN \& non-ECN, labelled A \& B) competing over each AQM. However, here there is 
not just one flow of each type, but 0--10 of each, as specified along the X-axis of \autoref{fig:l40r10}. In the second row, we also show results using Reno in place of CUBIC, which illustrates that there is no significant difference. 

The Y-axis shows the rate per flow normalized relative to the `fair' rate, meaning the link rate divided by the total number of flows. \autoref{fig:l40r10} shows results over 10\,ms base RTT and 40\,Mb/s capacity, which are representative of our other tests (not shown) over the 5 link rates and 5 RTTs used in \autoref{fig:1-1_aqm}.

\paragraph{Results} In the PIE and FQ-CoDel cases, all the flows have similar CCs, which results in near equal \textbf{per-flow rate}. The FQ-CoDel scheduler keeps the percentile range very tight. Except we sometimes see lower 1\%-ile (P1), when a hash collision maps more than one flow into the same queue.

DualPI2 ensures no flow gets much below the `fair' rate. When many Prague flows compete with few CUBIC flow (A8-B2, A9-B1), each Prague flow gets a little less, and the few CUBIC flows then consume the extra capacity. We traced this to reduced congestion signalling applied to the Classic flows because the weighted scheduler gives Classic traffic priority over approx.\ 10\% of the capacity.\footnote{Also, Prague gets additional marking when L4S queue delay goes above the 1\,ms threshhold, which leaves Classic traffic with an even greater link share than the scheduler allocates.} This confirms 
that the simple squared coupling of the DualPI2 AQM adequately counterbalances the 
more aggressive response of Prague for many different numbers of flow-types, not just 1-to-1.

\subsection{Different RTTs}\label{expts-diff-RTTs}
\begin{figure*}[!ht]
    \centering
    \begin{subfigure}[b]{\textwidth}
        \centering
        \includegraphics[width=\linewidth,trim={9cm 0 5cm 0},clip]{images/paper_plots/aqm_label.png}
    \end{subfigure}%

    \begin{subfigure}[b]{\textwidth}
        \centering
        \includegraphics[width=\textwidth,trim={1cm 19cm 2.7cm 0},clip]{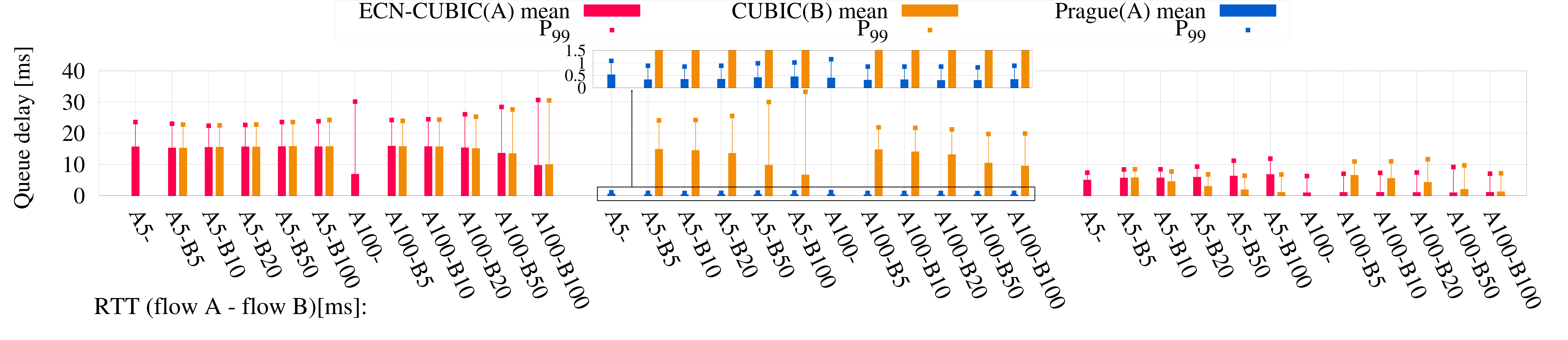}
    \end{subfigure}%

    \begin{subfigure}[b]{\textwidth}
        \centering
        \includegraphics[width=\textwidth,trim={1.4cm 3cm 1.5cm 4.5cm},clip]{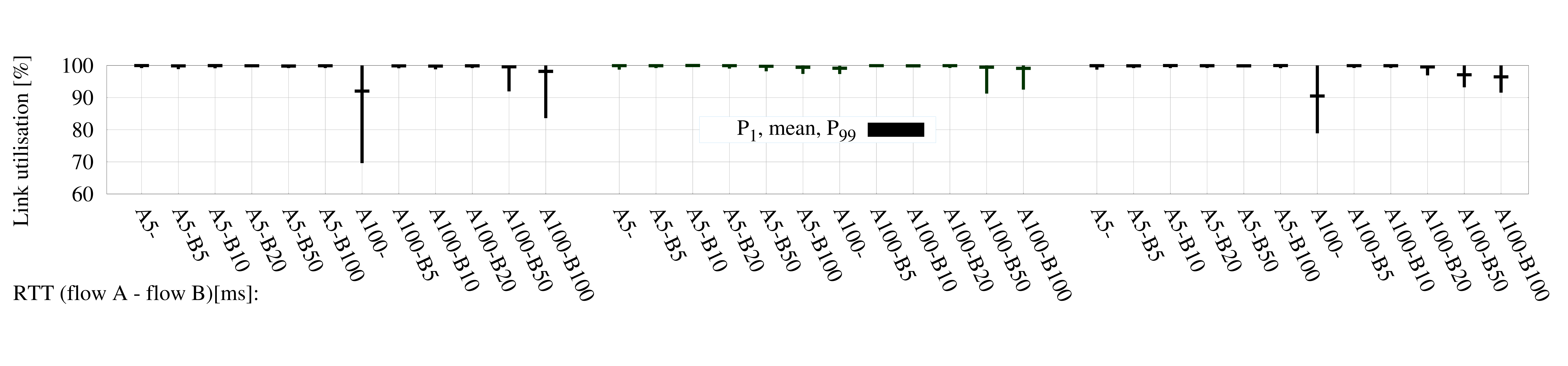}
    \end{subfigure}%

    \begin{subfigure}[b]{\textwidth}
        \centering
        \includegraphics[width=0.95\linewidth,trim={-2cm 0 0 0},clip]{images/paper_plots/plots_extra/aqm_label}
    \end{subfigure}%

    \begin{subfigure}[b]{\textwidth}
        \centering
        \includegraphics[width=\textwidth,trim={1cm 16cm 2.7cm 5cm},clip]{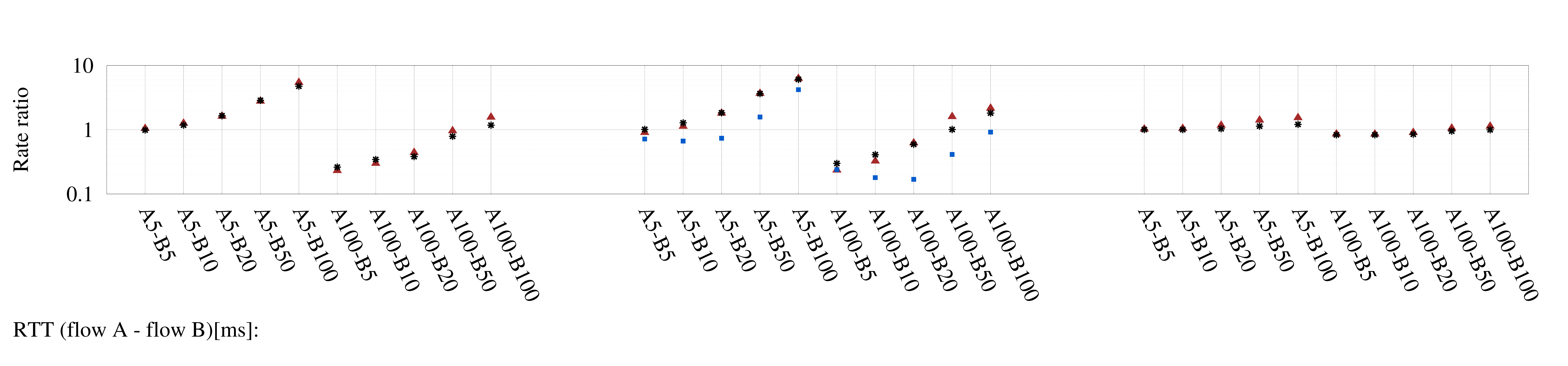}
    \end{subfigure}%

    \begin{subfigure}[b]{\textwidth}
        \centering
        \includegraphics[width=\textwidth,trim={1cm 3cm 2.7cm 3.9cm},clip]{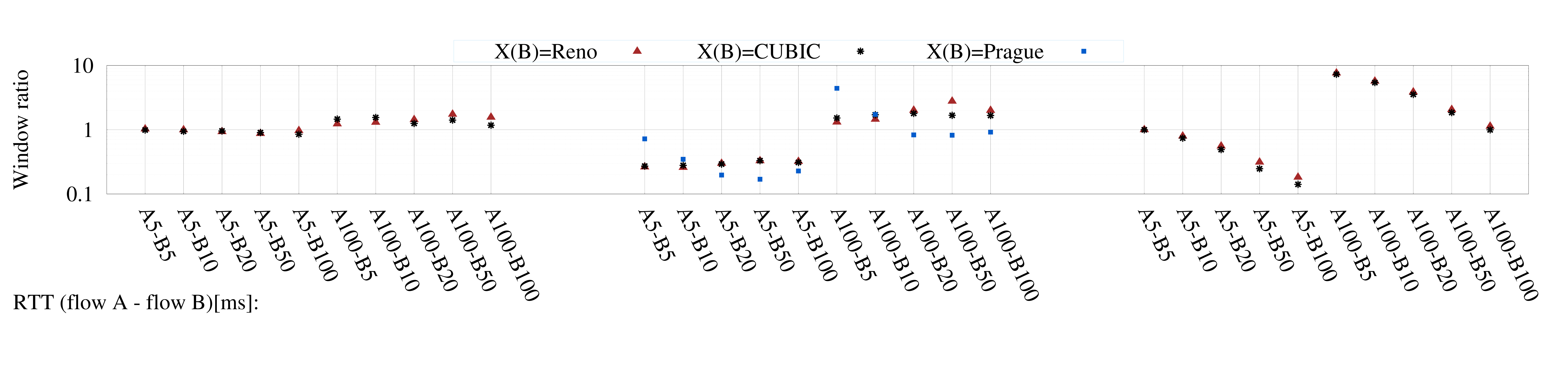}
    \end{subfigure}%

    \caption{Set-up as \autoref{fig:1-1_aqm} with 1 flow for each CC but mixed RTTs; Link: 40\,Mb/s.}
    \label{fig:mrtt2_link40}
    \vspace{-0.5cm}
\end{figure*}
\paragraph{Experimental setup} To evaluate the RTT-dependence of different CCs, we run the same pairs of CCs (labelled A or B) over the same AQMs as in the basic steady-state experiments, but with differing RTTs. For flow B, we also use Reno to check comparability with CUBIC, and Prague to check behaviour with itself. The base RTT of A is either 5 or 100\,ms and that of B is taken from the range 5, 10, 20, 50 and 100\,ms, as denoted along the X-axes of \autoref{fig:mrtt2_link40}. 
For example, 5-20 means 5\,ms for flow A and 20\,ms for B. Runs with no flow B are also included.  Results over a 40\,Mb/s link are shown, which are representative of tests run over the other 4 link rates (not shown).

\paragraph{Results} The zoomed inset in the top row of \autoref{fig:mrtt2_link40}, shows that even with longer RTT flows, Prague’s mean \textbf{queue delay} on DualPI2 is below 0.5\, ms for all tested cases, which is significantly lower than that achieved by Classic flows over PIE or FQ-CoDEL. Note that even Prague's P99 delay is always below 1\,ms, when it is driven by both the 1\,ms step threshold and coupled congestion marking. When running alone, its P99 is slightly above 1\,ms both for small and high RTTs, which is because its congestion marking is solely from the step threshold. The mean and P99 of DualPI2's Classic queue delay is lower or similar to PIE's. Indeed, as Classic RTT rises the mean and P99 diverge in the A5 cases over DualPI2 and the A100 cases over PIE. Both are due to CReno's larger sawteeth at high RTT, but only when there is no short RTT Classic flow to fill the valleys.{}
\bobk{Used to say ``, except for the A5-B100 case, where the P99 is a little higher than PIE's, but the mean is lower. This is due to significantly higher RTT of the Classic flow, which first overshoots and then recovers slowly after a congestion loss.''}

The second row shows that in steady state DualPI2 utilizes the link better than PIE or FQ-CoDel, with mean \textbf{utilization} close to 100\% in all cases --- even with a lone high RTT flow (A100-), where PIE and FQ-CoDel show significant underutilization. The P99 of DualPI2 falls to 90\% in the A100-B50 case, mainly due to rate variations of the Classic flow; and the A100-B100 case is similar. However neither of the Classic AQMs achieve better results in those cases. 

The third row shows that the scheduler of FQ-CoDel enforces near-perfect \textbf{rate balance}, as expected. Nonetheless, DualPI2 achieves similar rate balance to PIE, confirming that the effect of Prague’s RTT-independence algorithm is equivalent to that of a shared queue AQM on Classic flows. 
It is well-known that competing Classic TCP flows equalize their congestion windows, so that their rates become inversely proportional to their respective RTTs, evidenced by the \textbf{window ratio} of PIE staying around 1 (fourth row). For DualPI2, we see that when Prague has a lower RTT it reduces its window to keep the rate ratio within bounds. 

Additional rate ratio results are included for two competing Prague flows over different RTTs. Prague behaves as if its RTT is no lower than 25\,ms, so the rate ratio would be expected to flatten for RTTs below that. However, as RTT reduces further the rate ratio increases slightly. Further investigation pointed to an interaction between Prague's segmentation offloading and use of sojourn time in the AQM (see~\cite{Briscoe17b:sigqdyn_TR} for details).

We conclude so far that the very low queuing delay of Prague over DualPI2 is maintained with mixed RTTs, still without compromising utilization or rate balance.

\subsection{Dynamic Load}\label{expts-dynamic}

\paragraph{Experimental setup} To evaluate dynamic behaviour, we built on the basic steady-state experiments (\S\,\ref{expts-basic-steady-state}), but added emulated web traffic 
to each of the long-running flows between the two client-server pairs. Two web load profiles (high and low) were run, as defined in Appendix \ref{traffic_model}.

\begin{figure}[H]
	\centering
	\includegraphics[width=0.9\linewidth,trim={1.5cm 1cm 1cm 0},clip]{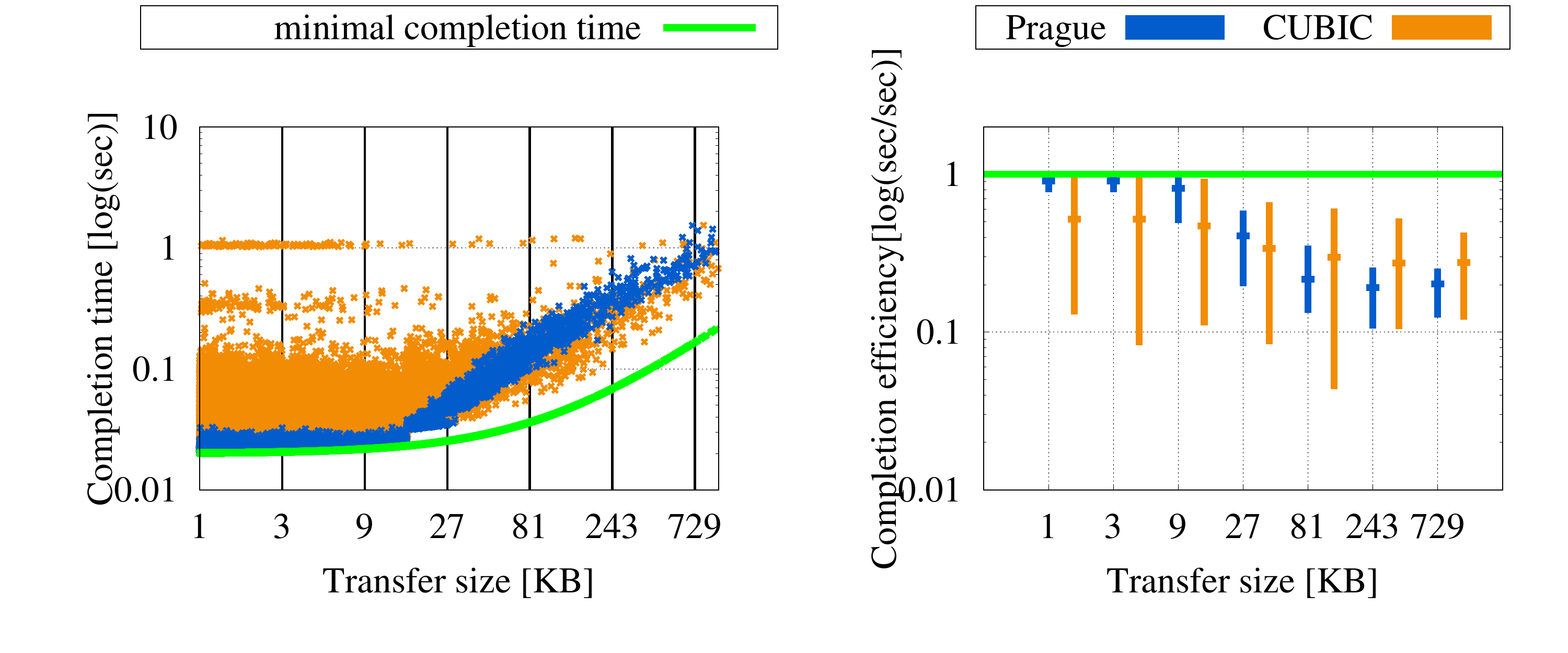}
	\caption{Efficiency representation (right) of example flow completion time scatter plot (left). High web load and 1 steady-state flow for each CC over DualPI2 AQM.}
	\label{fig:compl_expl}
	\vspace{-0.3cm}
\end{figure}%

As before, we show queuing delay and link utilization metrics, but we also show flow completion time (FCT) which is a more relevant metric than rate for short flows.
To better quantify the FCT distribution, we used
the Completion Efficiency representation on the right of \autoref{fig:compl_expl}.
Completion efficiency is defined as a theoretically achievable FCT (the green line at 1) divided by the actual FCT. 
The theoretically achievable FCT takes the RTT into account for the handshake but then downloads at full link speed. 
As one example, 
the left-hand side of \autoref{fig:compl_expl} shows the high web load DualPI2 AQM test case on a 40\,Mb/s
link with 10\,ms base RTT. It is a log-log scatter plot of the 
FCT to item size relation, and the green line along the bottom is the theoretically achievable FCT.
We
then binned the samples in log scale bins (base 3) and plotted the average,
\(1^\mathrm{st}\) and \(99^\mathrm{th}\) percentiles. 

\begin{figure*}[ht]
    \centering
    \begin{subfigure}[b]{\textwidth}
        \centering
    \includegraphics[width=\textwidth,trim={0 3cm 0 3.6cm},clip]{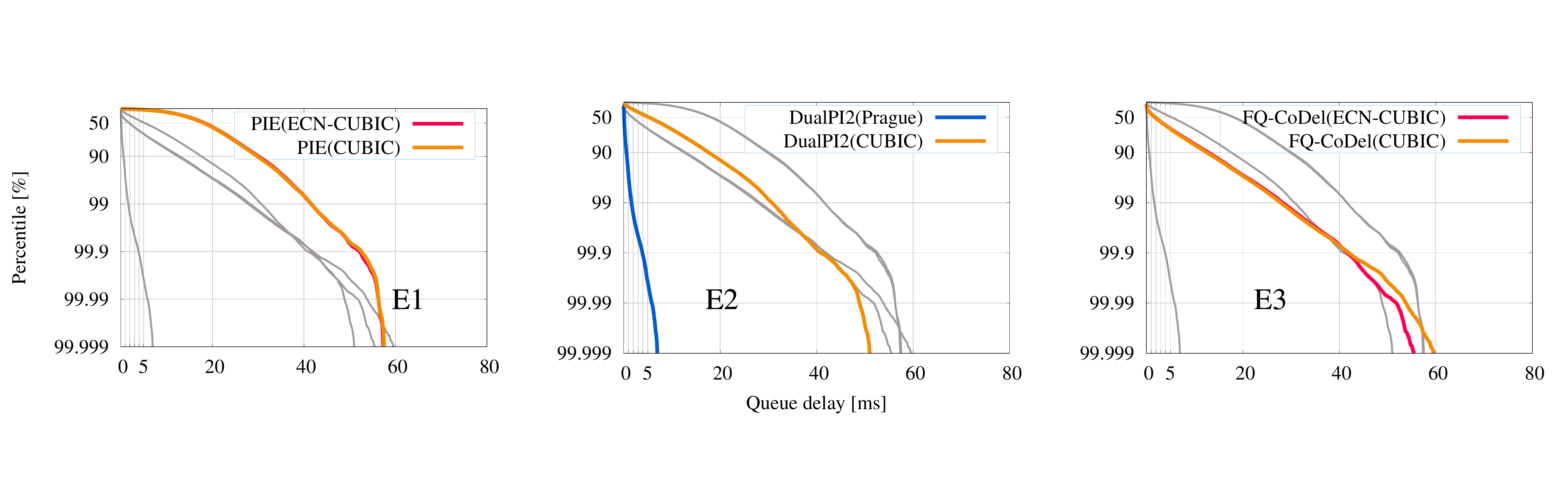}
    \end{subfigure}
	\caption{Log-scale complementary CDF comparing the DualPI2 AQM
	with PIE \& FQ-CoDel, each under the same challenging
	traffic scenario as \autoref{fig:1h-1h-ct}; Link: 120\,Mb/s, RTT: 10\,ms. Each experiment run  (E1,E2,E3) used 1 of the 3 AQMs and 2 CC types. The results for the AQM under test are shown in colour, with the other 2 AQMs in grey for comparison.}
    \label{fig:ccdf-delay-compare}
\end{figure*}

\begin{figure*}[!ht]
    \centering
    \begin{subfigure}[b]{\textwidth}
        \centering
        \includegraphics[width=\textwidth,trim={0 0 3.8cm 0},clip]{images/paper_plots/aqm_label.png}
    \end{subfigure}%

    \begin{subfigure}[b]{\textwidth}
        \centering
        \includegraphics[width=\textwidth,trim={1cm 3cm 3.6cm 4.5cm},clip]{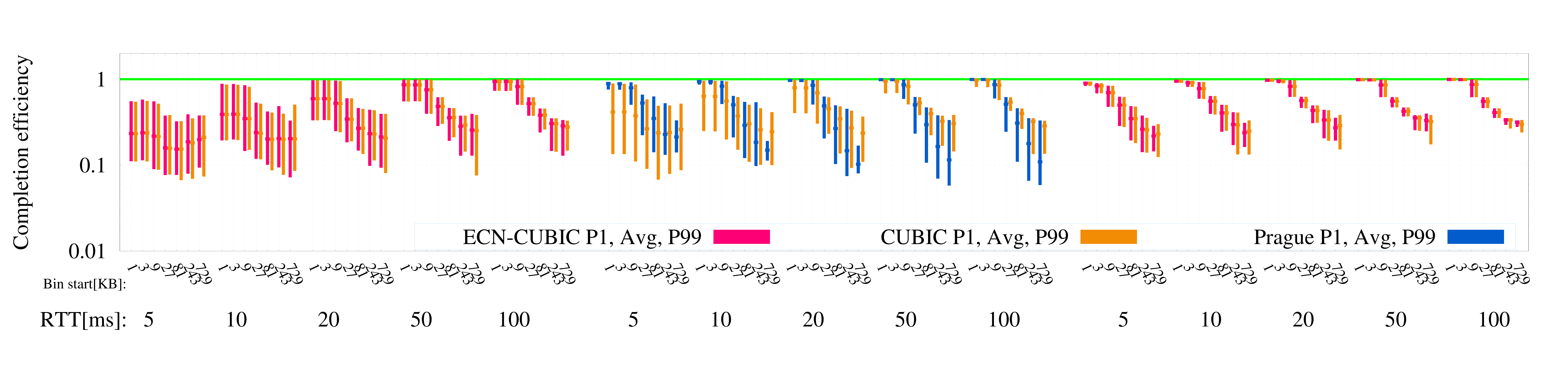}
    \end{subfigure}%

    \begin{subfigure}[b]{\textwidth}
        \centering
        \includegraphics[width=\textwidth,trim={1cm 3cm 3.5cm 3cm},clip]{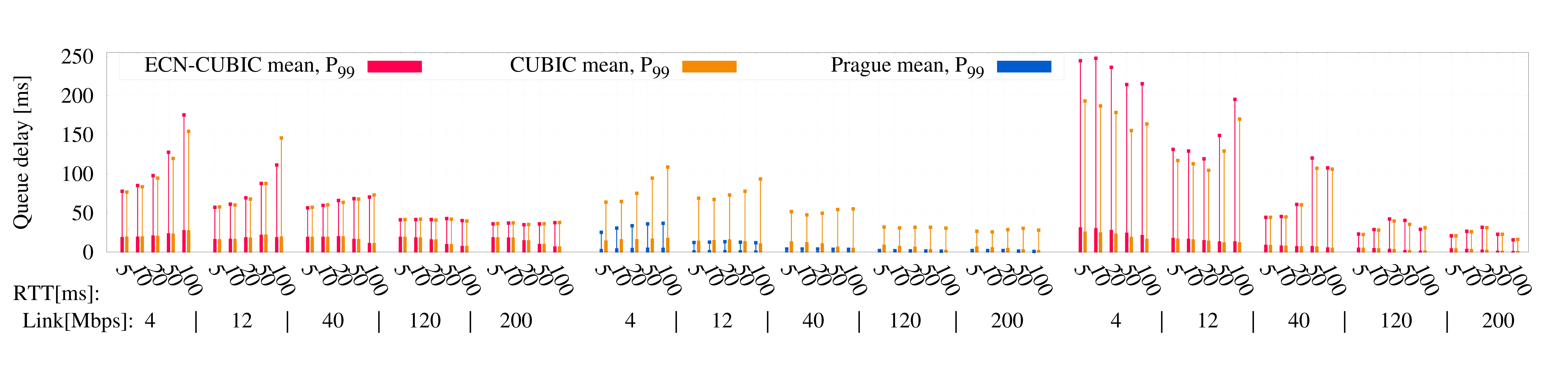}
    \end{subfigure}%

    \begin{subfigure}[b]{\textwidth}
        \centering
        \includegraphics[width=\textwidth,trim={1cm 3cm 3.3cm 3cm},clip]{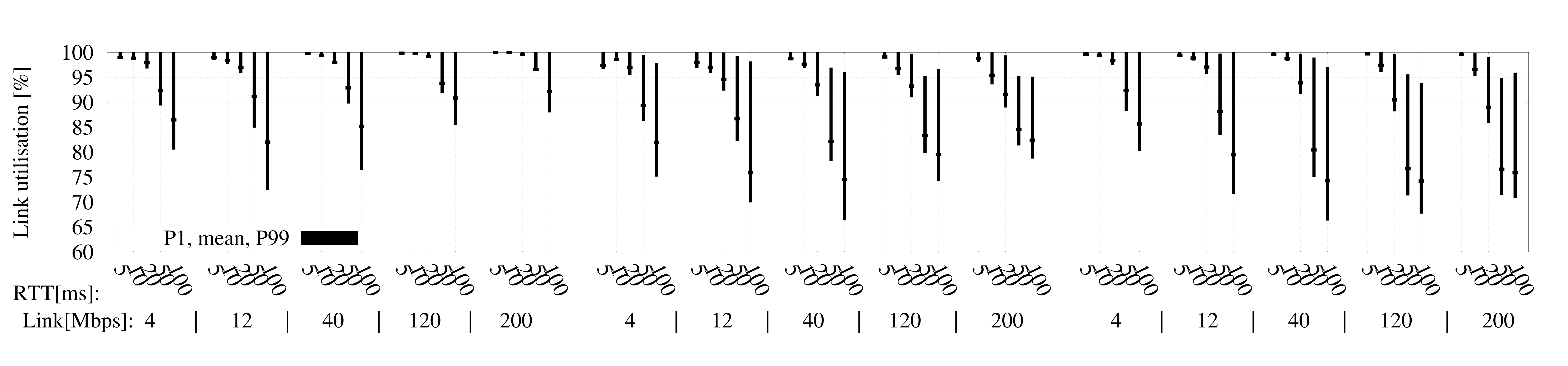}
    \end{subfigure}%
    \caption{Heavy dynamic workload. Top plot shows 
    	completion efficiency for a traffic mix of 1 long flow and 300 short requests per second for each CC; Link=120\,Mb/s, Base RTTs=10\,ms.
    Middle and bottom plots show queue delay and utilisation for the same (proportionately scaled) traffic mix but over the wider set of link/RTT combinations.}
    \label{fig:1h-1h-ct}
    \vspace{-0.4cm}
\end{figure*}

\autoref{fig:ccdf-delay-compare} and the top row of \autoref{fig:1h-1h-ct} show the most challenging high web load scenario over a 120\,Mb/s link. They are representative of the full set of experiments performed over all 5 link rates, which are summarized in the rest of \autoref{fig:1h-1h-ct}.

\paragraph{Results} The complementary CDFs of \textbf{queue delay} in \autoref{fig:ccdf-delay-compare} demonstrate that, even under heavy web load, L4S queue delay is below 1\,ms at P99 and an order of magnitude lower than PIE or FQ-CoDel at all the higher percentiles.

The \textbf{completion efficiency} results in \autoref{fig:1h-1h-ct} show an immediate latency benefit for short flows (\(<\)50KB), while many bigger flows (\(>\)50KB) suffer from Prague taking longer to get up to speed, in particular at RTTs above 50\,ms (see \S\,\ref{cc-reqs} later).

The mean and P1 of short CUBIC flows in both DualPI2 and PIE have poor completion efficiency. The two horizontal stripes at 1\,s and 0.3\,s on the left of \autoref{fig:compl_expl} largely explain this. They match the retransmission timeout of a TCP-SYN and a tail data packet respectively, which impact some flows randomly due to the prevailing loss environment. FQ-CoDel avoids this problem by prioritizing lone packets and Prague avoids it by supporting ECN-capable SYNs. CUBIC-ECN is not immune from SYN loss because the Classic ECN spec~\cite{rfc3168} does not allow TCP SYNs to be ECN-capable.

With dynamic load, Prague over DualPI2 preserves very low \textbf{queue delay} by sacrificing \textbf{link utilization}. 
\bob{Add if space. I think we have space to add here that DualPI2's tradeoff is good for interactive real-time, but do you think it would be closer to opinion than fact?}
FQ-CoDel's link utilization profile is similarly reduced, but without the low P99 queue delay, particularly for lower link rates (\(<\)120\,Mb/s), which is due to its short flow prioritization and RR scheduling. Low L4S delay is not at the expense of Classic delay, evidenced by DualPI2's Classic queue delay (mean and P99) often being significantly lower than FQ-CoDel \& PIE. 

We conclude that in general, high levels of dynamic traffic do not noticeably compromise either L4S or Classic queue delay with DualPI2.
Improvement to Prague's mechanisms for getting up to speed are discussed in \S\,\ref{cc-reqs} and its references, but evaluating them is outside our AQM-only scope.

\subsection{Unresponsive Flows and Overload}\label{expts-overload}

\begin{figure}[!ht]
\centering 
\begin{subfigure}[b]{\columnwidth}
    \centering
	  \includegraphics[width=\columnwidth,trim={1cm 6.2cm 3cm 0},clip]{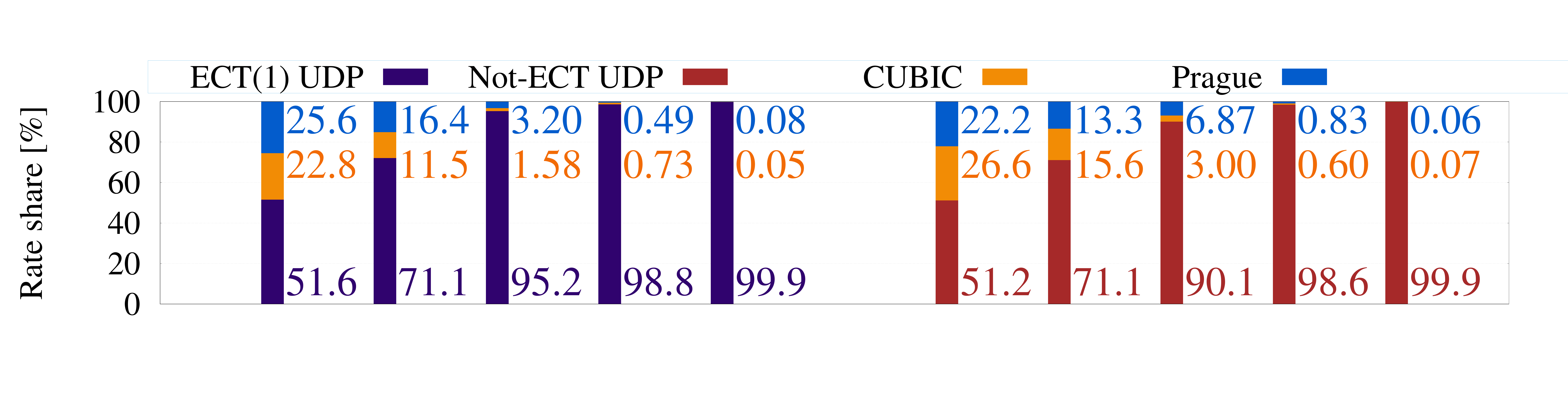}
\end{subfigure}
 \vspace{-0.01cm}

\begin{subfigure}[b]{\columnwidth}
    \centering
    \includegraphics[width=\columnwidth,trim={0.7cm 5.9cm 3cm 0.1cm},clip]{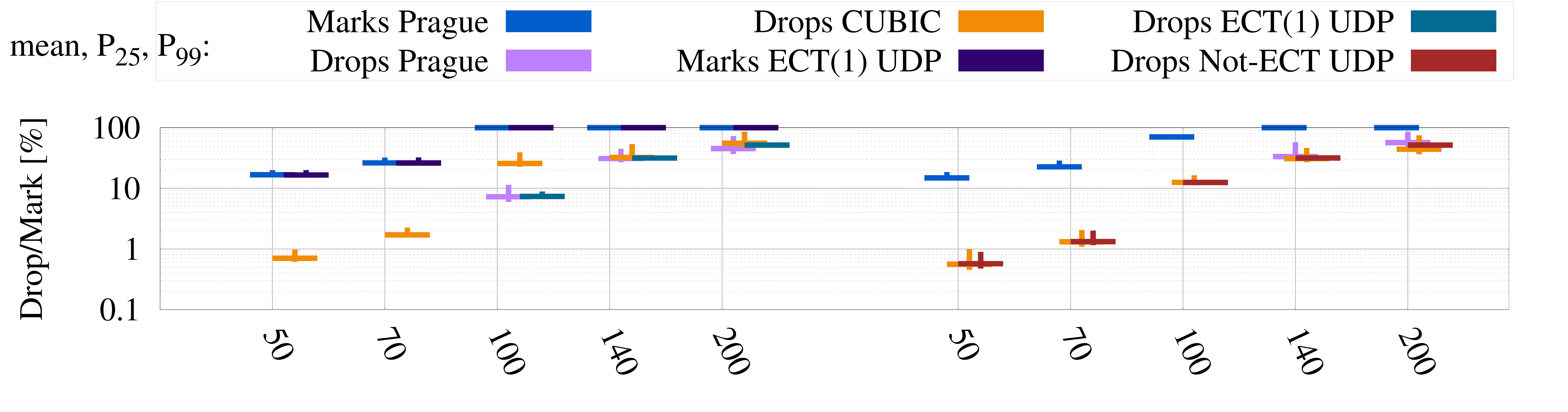}
\end{subfigure}
 \vspace{-0.3cm}

\begin{subfigure}[b]{\columnwidth}
    \centering
    \includegraphics[width=\columnwidth,trim={0 2cm 3cm 1.5cm},clip]{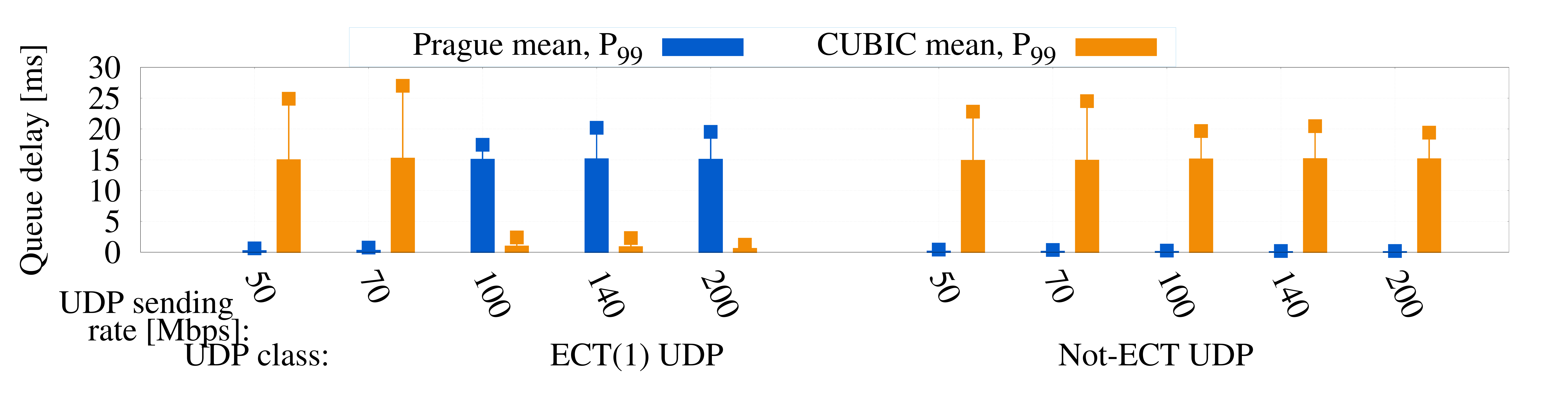}
\end{subfigure}

\caption{Overload experiments on a 100\,Mb/s link: 1 UDP flow + 10 long running TCP flows (5 flows for each CC). The plots compare a UDP flow classifying itself as ECT(1) and as not-ECT (L4S and Classic respectively).}
        \label{fig:overload}
\end{figure}

\paragraph{Experimental setup} 
To trigger overload, we added an unresponsive UDP flow
to 5 long-running flows of each CC type (Prague \& CUBIC) with 10\,ms base RTT over 
DualPI2 on a 100\,Mb/s bottleneck link. We tested 
5 different nominal UDP sending rates from 50\% to 200\% of link capacity, as shown on the X-axis of \autoref{fig:overload}, which also shows that we ran 2 sets of 
tests with the UDP flow's class marked as either ECT(1) (L4S) or not-ECT (Classic).
{} 

\paragraph {Results} The top row of \autoref{fig:overload} shows that, if a UDP flow targets the L4S queue rather than Classic, it hardly increases its \textbf{rate share}. Recall from \S\,\ref{overload} that this satisfies the design goal of not amplifying any existing attack, without resorting to flow rate policing, which is nonetheless still available as an orthogonal policy choice.

All three rows show that, with UDP rate lower than the bottleneck capacity (\(<\)100Mb/s), the choice of UDP class has no significant effect on any of the metrics of the responsive flows: \textbf{queue delay}, \textbf{marking probability} or \textbf{rate}. {}

If a UDP flow targets the L4S queue at \(\ge\)100\,Mb/s, it can increase L4S \textbf{queue delay} to the Classic target (15\,ms). Making the two queues behave as one on overload was a deliberate design choice (see \S\,\ref{overload}) to ensure that, if L4S traffic overloads the link, it cannot gain any latency benefit. For alternative choices see~\cite{Briscoe15e:DualQ-Coupled-AQM_ID}.

{}

We conclude that under overload the DualPI2 uses drop to shed load, effectively becoming a single queue Classic AQM, which excludes excessive unresponsive traffic from the advantages of both ECN and low delay.

\section{Deployment Considerations}\label{Deployment}

\subsection{Deployment Scenarios}\label{deployment}

Given access networks are invariably designed to bottleneck in one known
location, the DualQ does not have to be deployed in every buffer. Most of the
benefit can be gained by deployment at the downstream queue into the access
link, and home gateway deployment would address the upstream. Nonetheless, the DualQ is simple enough to be deployed at any potential bottleneck---in data centres, \iftr\else{}peerings, \fi{}access links or within end systems.

Also, the DualQ makes uncoordinated deployment of DCTCP practical in data centres, e.g.\ across
multi-tenant data centres or across community of interest networks connecting
private data centres. This extends the applicability of DCTCP beyond private networks where a centralized admin can coordinate deployment on a flag-day.

%

\subsection{Standardization Requirements}%
\label{standards}\label{codepoints}

The IETF has taken on L4S standardization work~\cite{Briscoe16a:l4s-arch_ID}.
It considered the pros and cons of various
candidate identifiers for senders to indicate L4S support and found that none 
were without problems, but
proposed ECT(1) as the least worst~\cite{Briscoe15f:ecn-l4s-id_ID}. As a consequence, the IETF has updated the
ECN standard at the IP layer (v4 and v6) to make the ECT(1) codepoint
available for experimentation~\cite{Black18:ecn-expts}.

The main issue is that there is only one spare codepoint. So, if senders use it to 
distinguish L and C packets, there is only one Congestion Experienced (CE) codepoint for the network to mark both L \& C packets (the `CE ambiguity problem'). 
CE is not ambiguous to a sending host, because it knows whether it was
sending L or C packets. However, it is ambiguous within the network, which has the following implications.

In the (unusual) case of multiple ECN bottlenecks along one path, a CE
mark from the first AQM is ambiguous to the second---it will not know whether it
was originally L or C. Nonetheless, it is benign for an L4S AQM to
classify any arriving CE packets into its L queue, because occasional early
packets do not trigger spurious retransmissions unless the five unlikely
conditions listed in Appendix B of \cite{Briscoe15f:ecn-l4s-id_ID} are all true.

Also, the `Classic' ECN standard~\cite{rfc3168} requires the network to treat ECT(1) and ECT(0) equally. So an L4S
host sending an ECT(1) packet does not explicitly know whether any resulting
CE-marking is from an L4S AQM or from a Classic AQM that doesn't understand L4S. If the L4S sender wrongly applies a small L4S
congestion response when the mark is actually from a Classic AQM in a shared queue, and
it is shared by competing Classic flow(s), the L4S flow will start to take an
`unfair' share of the capacity: the `coexistence problem'.

This important subject requires a whole paper, and it is beyond
the DualQ scope of the present paper anyway. In brief, it is widely believed
that most or perhaps all deployments of AQMs that support Classic ECN also use per-flow queuing (FQ). If so, there is only a coexistence problem in the unintended cases where more than one flow shares a per-flow queue, e.g.\ hash collisions or L3 VPNs. For such cases, and for any cases of single-queue Classic ECN AQMs, an approach has been
proposed~\cite{Briscoe19d:ecn-fallback} where the L4S sender detects which type
of AQM is marking the packets. The initial detection algorithm resulted in too many false positives, so the authors suggested
various avenues for improvement. Another
approach was proposed where senders proceed with L4S deployment while
monitoring for these potential problems, then targeted out-of-band testing could check for false positives~\cite{Briscoe15f:ecn-l4s-id_ID}.

Having chosen the new ECT(1) identifier, the IETF has also had to define its semantics. 
The 
square relationship between an L4S mark and a drop in this paper (Eqn.
(\ref{equ:mixcpl2a})) has been proposed for experimental
standardization~\cite{Briscoe15f:ecn-l4s-id_ID}.

The IETF has not specified a particular DualQ coupled AQM, but it has
specified the necessary relationship between AQMs in a DualQ structure in
order to couple them~\cite{Briscoe15e:DualQ-Coupled-AQM_ID}, so that multiple
implementations can be built, tested and compared, possibly using different
base AQMs internally. It has been proposed to recommend rather than standardize
a value for the coupling factor, \(k\), given differences would not prevent
interoperability.

The coupled DualQ AQM structure has already been
adopted by the cable industry as part of the mandatory low latency support added to
DOCSIS 3.1~\cite{CableLabs:DOCSIS3.1}.  

%
\section{Further Work}\label{Further Work}

\subsection{Validation}\label{Validation}

The results in this paper are largely similar to those published in 2015 \& 2019 as
technical reports~\cite{DeSchepper15b:DCttH_TR, DeSchepper19a:DCttH}. Subsequently, the results
regarding capacity sharing and latency isolation were validated
independently~\cite{BoruOljira20:L4S_validate} once it was clarified that L4S
sources need to employ pacing.

Heist \emph{et al}~\cite{Heist20:L4S_tests} identifies a number of scenarios
where L4S performance appears to suffer and raises a question about the impact
that bursty traffic can have on that performance. This question is
relevant to the AQM in this paper and it is believed that performance in the
presence of bursty traffic can be improved by basing
marking on the time it would take to drain the backlog at dequeue~\cite{Briscoe17b:sigqdyn_TR}, rather than at enqueue, which is what sojourn time measures. Work is ongoing
to identify which of the other issues are specific to L4S. Of those that are,
most are related to congestion control (see \S\,\ref{cc-reqs} below) and some have already been fixed after
having been reproduced using ns-3 models of L4S
components~\cite{Henderson20:L4S-eval}.

{}

\subsection{Congestion Control Roadmap}\label{cc-reqs}

TCP Prague~\cite{Briscoe21b:PragueCC-ID} as used in this paper is a set of fixes to issues in DCTCP so that it can act as a reference scalable
congestion control for proving and testing the L4S architecture. It is not
necessarily intended to be maintained as a production congestion control that is
optimized for the range of environments found on the Internet, although the open
source code is available for others to incorporate in parts or as a whole.
Nonetheless, it is sufficient to exercise the parameter space of our experiments
in order to evaluate the network mechanism of L4S, without which end-system
performance improvements would be moot. To further clarify the status of the
Prague Congestion Control, pending improvements are listed
below in priority order. They are a summary of the L4S transport 
layer behaviours identified by the IETF~\cite[Appx.\ A]{Briscoe15f:ecn-l4s-id_ID}, 
which are in turn adapted from the ``Prague L4S
requirements'', named after the meeting in Prague of a large group of DCTCP
developers that informally agreed them~\cite{Briscoe15:tcpPrague_Launch}. 
\begin{enumerate}[nosep]
	\item Less drastic exit from slow-start, similar goal to Flow-Aware
	(FA-DCTCP)~\cite{Joy15:FA-DCTCP} or 
	Paced Chirping~\iftr\else{\cite{Misund19a:Paced_Chirping_Linux}}\fi{};
	\item Faster-than-additive increase, 
	e.g.\ Adaptive Acceleration (A\(^2\)DTCP)~\cite{Zhang15:A2DTCP} 
	or Paced Chirping~\cite{Misund19a:Paced_Chirping_Linux};
	\item Yield fast to excessive delay increase, e.g.\ due to capacity reduction;
	\item Handle a window of less than 2, rather than grow the queue if base RTT is low~\cite[\S\,3.1.6]{Briscoe19a:TCP_Prague_Linux};
	\item Fall back to a Classic congestion response if a classic ECN bottleneck is
	detected (Prague contains an optional initial algorithm that was disabled for
	the present experiments~\cite{Briscoe19d:ecn-fallback});
	\item Improve RTT-independence of rate (see \S\,\ref{coupled} \& \cite{Briscoe21b:PragueCC-ID}).
\end{enumerate}

\section{Related Work}\label{related}

In 2002, Gibbens and Kelly~\cite{Gibbens02:Marking_priority} developed a scheme
to mark ECN in a priority queue based on the combined length of both queues.
However, they were not trying to serve different congestion controllers as in
the present work. In 2005 Kuzmanovic~\cite[\S5]{Kuzmanovic05:ECN_SYN_ACK}
presaged the main elements of DCTCP showing that ECN should enable a na\"{\i}ve
unsmoothed threshold marking scheme to outperform sophisticated AQMs like the
proportional integral (PI) controller. It assumed smoothing at the sender, as
earlier proposed by Floyd~\cite{Floyd94:ECN}.
 
Wu et al.~\cite{Wu:2012:TED:2413176.2413181} investigates a way to incrementally
deploy DCTCP within data centres, marking ECN when the temporal queue exceeds a
shallow threshold but using standard ECN~\cite{rfc3168} on end-systems. Also in a data centre context, Irteza \emph{et al.}~\cite{Irteza14:dc-cc-coexist} investigates potential pre-existing solutions to the congestion control coexistence problem and finds them unsatisfactory.
Kuhlewind \emph{et al.}~\cite{7063495} showed that DCTCP and Reno could co-exist in the
same queue configured with a form of WRED~\cite{Clark98:RIO} classifying on ECN
rather than Diffserv, but only over a limited set of conditions. Judd~\cite{Judd15:DCTCP_Pitfalls} uses Diffserv scheduling to
partition data centre switches between DCTCP and classic traffic in a financial
data centre scenario, but as already explained this relies on management
configuration based on prediction of the traffic matrix and its dynamics, which
becomes hard on low stat-mux links. Fair Low Latency (FaLL)~\cite{Xue15:FaLL} is
an AQM for DC switches building on CoDel{}. Unlike the
DualQ, FaLL inspects the transport layer of sample packets to focus more marking
onto faster flows while keeping the queue short.

\section{Conclusion}\label{conc}

Classic TCP induces two impairments: queuing delay and loss. A good AQM can
reduce queuing delay but then TCP induces higher loss. In a low stat-mux
link, there is a limit to how much an AQM can reduce queuing delay without TCP's
sawteeth introducing a third impairment: under-utilization. Thus TCP is like a
balloon: when the network squeezes one impairment, another bulges out.

This paper moves on from debating how the network should best squeeze the TCP
balloon. It recognizes that the problem is now wholly outside the network:
Classic TCP (the balloon itself) is the problem. But this does not mean the
solution is also wholly outside the network. This paper has shown that the
network plays a crucial role in enabling hosts to transition away from the
Classic TCP balloon. The `DualQ Coupled AQM' detailed in this paper is not
notable as somehow a `better' AQM than others. Rather, it is notable as a
coupling between two AQMs in two queues---as a transition mechanism to enable
hosts to dispense with their old TCP balloon.

Hosts will then be able to transition to a member of the family of scalable
congestion controls. This can still be likened to a balloon. But it is a tiny
balloon (near-zero impairments) and, importantly, it will stay the same tiny
size (invariant impairments as BDP scales). Whereas the Classic TCP balloon is
continuing to inflate (worsening impairments) as BDP scales. This transition to
a scalable regime is as important as the low delay that L4S offers today.

The paper provides not just the mechanism but also the incentive for
transition---the tiny size of all the impairments. For link rates from
4--200\,Mb/s and RTTs from 5--100\,ms, our extensive testbed experiments with a
wide range of heavy load scenarios have shown near-zero congestion loss;
sub-millisecond average queuing delay (roughly 500\,\(\mu\)s) with tight
variance (P99 of 2\,ms); and near-full utilization.

We have been careful as far as possible to do no harm to those still using the
Classic service. Also, given the network splits traffic into two queues, when it
merges them back together, we have taken great care not to enforce
flow `fairness'. Nonetheless, if hosts are aiming for flow `fairness' they will
get it, while remaining oblivious to the difference between Scalable and Classic
congestion controls.

We have been careful to handle overload in the same principled way as normal
operation, preserving the same very low delay for L4S packets, and dropping
excess load as if the two queues were one.

And finally, we have been careful to heed the zero-config requirement of recent
AQM research, not only inherently auto-tuning AQMs to link rate, but
also shifting RTT-dependent smoothing to end-systems, which know their own RTT.

\bibliographystyle{acm}
\bibliography{dctth}


%
\clearpage
\appendix
\subsection{DualQ Coupling Factor}\label{coupled-aqms}

Here we derive the constant coupling factor \(k\) defined as follows in \autoref{equ:mixcpl2a} of \S\,\ref{coupled} using the terminology defined there:
\begin{equation}
	k = \sqrt{\frac{8}{3}} \left.\frac{R_C}{f(R_L)}\right|_{R_b^*}
\end{equation}

\(R_C\) is the Classic RTT averaged over the Classic sawteeth. So it can be related geometrically to the RTT at the AQM's operating point:
\begin{equation}\label{eqn:reno_R_C}
	R_C \approx \frac{(1+\beta_C)}{2}(R_b+\mathrm{target}),
\end{equation}
where \(\beta_C\) is the multiplicative decrease factor of the Classic flow (0.5 for Reno or 0.7 for CReno). (\(R_b+\mathrm{target}\)) is roughly where the peaks of the Classic RTT sawteeth settle, not their average~\cite[\S\,3.3]{Briscoe21c:pi2param}.\footnote{This is because, at the scale of today's Internet, the duration of each sawtooth of a Classic congestion control (the recovery time) is usually significantly longer than the time it takes for a Classic AQM like PI\(^2\) to converge to a stable congestion level.} So it is factored down to the average by \((1+\beta_C)/2\).

To calculate \(k\) it is only necessary to know the value of \(f(R_L)\) at the reference RTT, \(R_b^*\), chosen by the operator. It is likely that this will be an RTT where the implementer has ensured that \(f(R_L)\approx R_L\), and one can assume that \(R_L \approx R_b\) because the L queue is negligible. For instance, for the current Linux implementation of Prague \(f(R_L) = \mathrm{max}(R_L, 25\,\mathrm{ms})\). So, if the operator chooses \(R_b^* = 25\)\,ms, which is a typical base RTT for Internet traffic~\cite{Briscoe21c:pi2param}, it is indeed true there that \(f(R_L)\approx R_b\). In such cases:
\begin{align}
	k \approx \sqrt{\frac{8}{3}} \frac{(1+\beta_C)}{2}\left(1+\frac{\mathrm{target}}{R_b^*}\right),
\end{align}

Then, taking the recommended value of \(\mathrm{target} =15\,\)ms, \(k=1.96\) for Reno%
\footnote{\autoref{eqn:reno_R_C} is only valid for \(R_b\leq \mathrm{target}*\beta_C/(1-\beta_C)\). For base RTTs above this, the window sawtooth under-utilizes the buffer. Therefore the values plugged in are outside the range of validity for Reno, but inside for CReno. However, the approximation is still close enough in practice, particularly because the tips of the sawteeth actually settle slightly above the AQM target, which slightly expands the range of validity.}
or 2.22 for CReno. 

\subsection{Evaluation metrics}\label{metrics}

Each experiment (lasting 250\,s) was performed with a specified TCP variant
configured on each client-server pair A and B and a
specified AQM, bottleneck link speed and RTT on the AQM server. The measurements were started after long-runnning TCP flows reached their steady states, skipping the slow start. To skip the slow start, we added a waiting time of $T_w$ seconds before starting the measurements, which was calculated as $T_w = 5 + \mathrm{rate} * \mathrm{rtt} / 100$. The slow start was not skipped for short dynamic flows.

Queuing delay was measured inside the qdiscs per packet, at dequeue time. All samples were further processed to derive mean and ${99}^{th}$ percentile.

Drop probability was also measured inside the qdiscs by keeping the counter of dropped packets, while mark probability was obtained by inspecting the packet headers and checking which packets were marked. Both mark and drop probability are represented as a percentage of all packets that passed the network interface where the qdisc was installed. All collected 1-second samples per experiment were processed to obtain mean, ${25}^{th}$ and ${99}^{th}$ percentiles.

To evaluate rate and window behaviour,
we measured throughput for each long-running flow by capturing all traffic at the AQM node. Since some experiments had an unequal number of flows for each CC, we sampled average rate per flow and normalized it by dividing by the fair rate per flow. Fair rate was calculated by dividing the total link capacity by total number of competing flows. The same approach was applied to window measurements. For convenience, Window size was approximated on the AQM node as $W = r* (qd_{avg} + R)$, where $r$ is measured rate in bytes, $qd_{avg}$ is average queue delay and $R$ is base RTT (both delay and RTT were converted to fractions of a second).
All rate and window measurements were sampled per second, and all samples were further processed to derive mean, ${1}^{st}$ and ${99}^{th}$ percentiles. Rate and window balance ratio was calculated by dividing average rate/window per flow of the ECN-capable CC by average rate/window per flow of the Classic CC for the entire experiment.

Utilization was measured by comparing all traffic captured at the interface to the total link capacity in 1-second intervals; as with other measurements, we derived mean, ${1}^{st}$ and ${99}^{th}$ percentiles by processing all obtained samples. 

For short dynamic flows, the client logged the completion time and downloaded size. Timing was 
started just before opening the TCP socket, and stopped after the connection close by the server was detected.

\subsection{Web Traffic Model}\label{traffic_model}

To model the web traffic in the dynamic load experiments (\S\,\ref{expts-dynamic}), 
an exponential arrival
process was used with an average of 1 (low load) or 10 (high load) requested items per
second for the 4\,Mb/s link capacity, scaled for the higher link speeds up to 50
(low) or 500 (high) requests for the 200\,Mb/s links. Every request opened a new
TCP connection, closed by the server after sending data with a size according to
a Pareto distribution with $\alpha=0.9$ and a minimum size of 1\,kB and maximum
1\,MB.
%


\end{document}